\documentclass[12pt, A4]{article}
\usepackage[top=1in,left=1in,bottom=1in,right=1in]{geometry}

\usepackage{sectsty}
\usepackage[hyphens]{url}
\usepackage[breaklinks]{hyperref}
\usepackage{amsmath, amssymb, amsthm, amsbsy, amsfonts, cases}
\usepackage{graphicx, comment, enumitem}
\usepackage{algorithm, algcompatible}
\usepackage[noend]{algpseudocode}
\sectionfont{\fontsize{16}{16}\selectfont}
\subsectionfont{\fontsize{14}{14}\selectfont}
\usepackage{bigints}
\usepackage[mathscr]{euscript}
\usepackage[makeroom]{cancel}
\usepackage{graphicx}
\usepackage[table]{xcolor}
\usepackage{subcaption}
\usepackage{enumerate, verbatim}
\usepackage[font={small,it,singlespacing}]{caption}
\usepackage{lscape}
\usepackage{caption}
\usepackage{hyperref}
\usepackage[sort, round]{natbib}
\usepackage{parskip}
\usepackage{array}
\usepackage{multirow}
\usepackage{graphicx}
\usepackage{wrapfig}
\usepackage{lscape}
\usepackage{rotating}
\usepackage{epstopdf}
\usepackage[compact]{titlesec}

\usepackage{color, colortbl, setspace}
\definecolor{Gray}{gray}{0.9}

\usepackage{boldline}
\hypersetup{colorlinks,citecolor=blue}
\graphicspath{{./images//}}

\newcommand{\bs}{\boldsymbol}
\newcommand{\R}{\mathcal{R}}
\newcommand{\T}{\mathcal{T}}
\newcommand{\x}{\mathbf{x}}

\newcommand{\s}{\mathbf{s}}

\newcommand{\vv}{\mathbf{v}}

\newcolumntype{L}[1]{>{\raggedright\let\newline\\\arraybackslash\hspace{0pt}}m{#1}}
\newcolumntype{C}[1]{>{\centering\let\newline\\\arraybackslash\hspace{0pt}}m{#1}}
\newcolumntype{R}[1]{>{\raggedleft\let\newline\\\arraybackslash\hspace{0pt}}m{#1}}

\theoremstyle{plain}
\newtheoremstyle{exampstyle}
  {6pt} 
  {\topsep} 
  {} 
  {} 
  {\bfseries} 
  {.} 
  {.5em} 
  {} 
\theoremstyle{exampstfyle}\newtheorem{defn}{Definition}
\theoremstyle{exampstyle}\newtheorem{lem}{Lemma}
\theoremstyle{exampstyle}
\theoremstyle{exampstyle}
\theoremstyle{exampstyle}
\theoremstyle{exampstyle}

\titlespacing*{\section}{0pt}{3pt}{3pt}
\titlespacing*{\subsection}{0pt}{3pt}{3pt}

\begin{document}
\title{\large{\textbf{Differentially Private Data Release via\\ Statistical Election to Partition Sequentially}}}
\author{\small{\textbf{Claire McKay Bowen, Fang Liu, Bingyue Su}}
\footnote{\noindent Claire McKay Bowen is the Lead Data Scientist for Privacy and Data Security at the Urban Institute. Fang Liu is Professor and Bingyue Su is a doctoral student in the Department of Applied and Computational Mathematics and Statistics, University of Notre Dame, Notre Dame, IN 46556. Claire McKay Bowen was supported by the National Science Foundation (NSF) Graduate Research Fellowship under Grant No. DGE-1313583 during part of the development of this paper. Fang Liu was supported by the NSF Grants  \#1546373, \#1717417, and the University of Notre Dame Faculty Research Support Initiation Grant Program. Bingyue Su is supported by the NSF Grant \#1717417.} 
\vspace{-3.5cm}
}
\date{}
\maketitle{}

\begin{abstract}
\noindent Differential Privacy (DP) formalizes privacy in mathematical terms and provides a robust concept for privacy protection. DIfferentially Private Data Synthesis (DIPS) techniques produce and release synthetic individual-level data in the DP framework. One key challenge to developing DIPS methods is preservation of the statistical utility of synthetic data, especially in high-dimensional settings. We propose a new DIPS approach, STatistical Election to Partition Sequentially (STEPS) that partitions data by attributes according to their importance ranks according to either a practical or statistical importance measure. STEPS aims to achieve better original information preservation for the attributes with higher importance ranks and produce thus more useful synthetic data overall. We present an algorithm to implement the STEPS procedure and employ the privacy budget composability to ensure the overall privacy cost is controlled at the pre-specified value. We apply the STEPS procedure to both simulated data and the 2000-2012 Current Population Survey youth voter data. The results suggest STEPS can better preserve the population-level information and the original information for some analyses compared to PrivBayes, a modified Uniform histogram approach, and the flat Laplace sanitizer.\\

\noindent \textit{\textbf{keywords}}:  privacy budget, DIfferentially Private Data Synthesis (DIPS), general utility, propensity score, universal histogram, hierarchical
\end{abstract}

\newpage
\setstretch{1.05}
\section{Introduction}\label{sec:intro}
Various approaches have been developed to provide protection for individual sensitive information when releasing data to researchers or the public. Among these approaches, data synthesis is a popular technique that generates synthetic individual-level data given the original data \citep{rubin1993discussion, little1993statistical, liu2003,raghunathan2003multiple, reiter2003,  liu2004, reiter2009, drechsler2011book}. The data synthesis concept is appealing as it provides surrogate data sets that have the same structure as the original data, and users may perform their own analyses as if they had the original data. A recent data synthesis approach is DIfferentially Private data Synthesis (DIPS) \citep{bowen2020comparative} that  performs data synthesis based on differential privacy (DP), a mathematical framework for controlling privacy risk with a prespecified privacy budget  \citep{dwork2006calibrating}. 

Some researchers consider DIPS methods as ``flat'' or one-step in the sense that DP noises are injected all at once or in parallel, such as the smooth and perturbed histograms \citep{wasserman2010statistical}, MOdel-based DIPS (MODIPS) \citep{liu2016model}, DPCopula \citep{DPcopula}, and PrivBayes \citep{PrivBayes},  among others.   There also exist DP procedures that partition data sequentially and inject random noises in a hierarchical manner to improve accuracy of certain types of queries in low-dimensional settings. Though the focus of these DP mechanisms is not data synthesis, they may be used for data syntheses with some adaptations or  with additional steps. We refer to this type of DIPS as sequential or hierarchical as opposed to flat. For example, \citet{xiao2010differentially} proposed Privelet via a two-step wavelet-based multidimensional partitioning approach to release range count queries. Privelet is effective in the one-dimensional case but makes only slight improvements in the two-dimensional case, and the performance could be even worse at higher dimensions \citep{qardaji2013geo}. \citet{xiao2012dpcube} presented DPCube as a two-phase partitioning approach for data cubes. \citet{gardner2013share} implemented DPCube in biomedical data to explore its practical feasibility on real-world data, but discovered that DPCube is inefficient in constructing accurate high-dimensional histograms. \citet{hay2010boosting} developed the universal histogram (UH) to inject noise to histogram bi-partitioning with improved accuracy in histogram bin counts by exploring the inherent consistency constraints. \cite{qardaji2013understanding} extended the method to relatively high dimensional data. \citet{hay2016principled} conducted an extensive comparison on most of the above mentioned algorithms using a set of evaluation principles (DPBench), and provided valuable insights on the pros and cons on the  methods for answering 1- and 2-dimensional range queries. \citet{li2017partitioning} introduced the privacy-aware partitioning mechanism and the utility-based partitioning mechanism that depend on a public but personalized privacy parameter. These methods have not been implemented to real-world data likely due to being ``unable to provide corresponding error guarantees with such procedures for general functions'' \citep{cummings2018individual}. 

In summary, most of the hierarchical procedures focus on low dimensional data with numerical attributes. To explore the potentials of hierarchical sanitization in improving the utility of synthetic data in relatively high dimensional settings when the original data have both categorical and numerical attributes, we propose a new DIPS procedure -- \textbf{ST}atistical \textbf{E}lection to \textbf{P}artition \textbf{S}equentially (STEPS). STEPS injects noises to a hierarchical histogram. The structure of the histogram is informed either by domain or prior knowledge, or by a statistical metric that explores the inherent statistical information in the original data. Different layers on the same branch in the constructed hierarchical histogram contains different attributes. The hierarchical histogram built by STEPS is subject to equality constraints among the nodes from different layers on the same branch; thus a post-processing procedure similar to the UH approach \citep{hay2010boosting} is used to ensure that the equality constraints are satisfied. The final step of STEPS is to generate synthetic data from the differentially private hierarchical histogram. We expect that STEPS will improve the statistical utility of the released data compared to data synthesized through a hierarchical histogram with random partitioning given that the data partitioning in STEPS is an informed decision, leveraging knowledge in the data or publicly available relevant knowledge. We apply STEPS to simulated data and the youth voter data from the 2000-2012 Current Population Survey (CPS) and benchmark its utility against  some DIPS approaches. To assess the overall utility of  the synthetic data, we also develop a propensity score based method that provides a general utility metric and a holistic measure for the similarity between two data sets of the same structure. 

The remainder of the paper is organized as follows. Section \ref{sec:methodology} provides some preliminaries regarding DP and introduces the STEPS procedure. Section \ref{sec:simulation} applies STEPS to simulated data and compares its capability in preserving the population-level information against PrivBayes and a modified UH procedure. Section \ref{sec:voter} implements the STEPS method to the CPS youth voter data and compares the statistical utility of the synthetic data generated by STEPS, PrivBayes, a modified UH procedure, and a flat Laplace sanitizer via several utility analysis. In Section \ref{sec:disc}, we discuss the implications of our results and provide future research directions.


\section{Methodology}\label{sec:methodology}
\subsection{Preliminaries on Differential Privacy}
DP provides a mathematical and rigorous framework for protecting individual information in a data set, regardless of the background knowledge or behaviors of data intruders, when releasing queries to the public. Query results, in statistical terminology, are statistics; so we use queries, query results, and statistics, interchangeably in this discuss and denote them by $\s$. We denote the data for privacy protection by $\x=\{x_{ij}\}$ for $i=1,\ldots,n;j=1,\ldots,p$. Each row  $\x_i$ represents an individual record with $p$ variables/attributes. We assume that the sample size $n$ and the number of attributes $p$ are public knowledge and carry no privacy information.
\begin{defn}[\textbf{Differential Privacy} \citep{dwork2006calibrating}]\label{def:dp}
Let $d(\x,\x')=1$ represents all possible ways that data set $\x'$ differing from $\x$ by one individual. A sanitization algorithm $\R$ gives $\epsilon$-DP, if for all data sets $(\x,\x')$ that is $d(\x,\x')=1$ and all result sets $Q\subseteq \T$, where  $\T$ denotes the output range of $\R$,  to queries/statistics $\s$ that
\begin{equation}\label{eqn:dp}
\left|\log\left(\frac{\Pr(\R( \s(\x)) \in Q)}{\Pr(\R( \s(\x'))\in Q)} \right)\right|\le\epsilon,
\end{equation}
\noindent where  $\epsilon>0$ is the privacy loss parameter.
\end{defn}

The smaller $\epsilon$ is, the more noises are injected to the statistic $\s$ via the sanitized algorithm $\R$; and each individual in the data set has a lower risk of being identified or having their sensitive information disclosed, because $\s$ would be about the same with and without that individual in the data. In addition to the $\epsilon$-DP in Definition \ref{def:dp}, there are also  conceptual relaxations of the ``pure'' DP such as the approximate DP \citep{dwork2013algorithmic}, probabilistic DP \citep{machanavajjhala2008privacy}, and concentrated DP \citep{dwork2016concentrated}, among others, so to lessen the amount of noises injected by DP methods by sacrificing a certain amount of privacy.

In regards to what value of $\epsilon$ is considered appropriate or acceptable for practical use, \citet{dwork2008survey} states the choice of $\epsilon$ is a social question. \citet{abowd2015revisiting} acknowledge this and suggest $\epsilon$ at $0.01$ to $\ln(3)$, or even up to 3 in releasing certain statistics in social and economic studies. A wide range of $\epsilon$ values have been examined in the literature. For instance, \citet{machanavajjhala2008privacy} applied DP in the OnTheMap data (commuting patterns of the United States population) and used $(\epsilon=8.6,\delta=10^{-5})$-probabilistic DP (a relaxation of the pure DP) to synthesize commuter data. \citet{DPcube} and \citet{DPcopula} used $\epsilon=1$ in their experiments. These examples suggest there are many factors that affect the choice of $\epsilon$, including the type of information released to the public, social perception of privacy protection, statistical accuracy of the release data, among others. For a socially acceptable $\epsilon$ given a certain type of information, a differentially private mechanism should aim for maximizing the accuracy of released information. In other words, choosing an ``appropriate'' $\epsilon$ is essentially a question of balancing the privacy loss and the accuracy of the released information.

An important property of DP is that the privacy cost increases for every new query released from the same data set, because more information is ``leaked'' with releasing more query results. Therefore, the data curator must track all statistics calculated on the data set to guarantee the privacy budget does not exceed the prespecified level. For example, if all $q$ queries are sent to data set $\x$, then $\epsilon/q$ can be allocated to each query to ensure the privacy budget is maintained at $\epsilon$ overall per the \textit{sequential composition} principle \citep{mcsherry2009privacy}. When no overlapping information is requested by different queries, such as when they are calculated from disjoint subsets of a data set, the privacy cost does not accumulate. In such a case, the \textit{parallel composition} principle applies and the overall privacy cost is the maximum privacy budget spent across all the queries \citep{mcsherry2009privacy}.

A common and easy way to implement DP is the Laplace mechanism. A key concept for the Laplace mechanism is the $l_1$ global sensitivity of statistic $\s$ (either a scalar or a vector) is $\Delta_1=\mbox{max}_{\x,\x', d(\x,\x')=1} \|\s(\x)-\s(\x')\|_1$ for all $d(\x,\x')=1$ \citep{dwork2006calibrating}. Global sensitivity can also be defined in other forms, such as the $l_2$ global sensitivity \citep{dwork2013algorithmic} and $l_p$ global sensitivity for any $p\ge1$ \citep{liu2016generalized}.

\begin{defn}[\textbf{Laplace mechanism} \citep{dwork2006calibrating}]\label{def:lap}
The Laplace sanitizer adds noise to statistics $\s=(s_1,\ldots,s_K)$ via $s_k^\ast=s_k+e_k$, where $e_k{\sim} \text{Lap}(0,\Delta_1/\epsilon)$ and is independent for $k=1,\ldots,K$ and $\Delta_1$ is the $l_1$ global sensitivity of $\s$.
\end{defn}
When $\epsilon$ is small or $\Delta_1$ is large, more Laplace noise is added to $\s$. Other common DP mechanisms include the Gaussian mechanism that is built upon the approximate or the probabilistic DP \citep{dwork2013algorithmic, liu2016generalized}, and the exponential mechanism that can release both numerical and non-numerical queries \citep{mcsherry2007mechanism}, among others. The definition of the exponential mechanism is presented below, which will be used in our proposed STEPS procedure.

\begin{defn}[\textbf{exponential mechanism} \citep{mcsherry2007mechanism}]\label{defn:exponential}
Let $u$ be a utility function that assigns a score to each possible output of a query. The exponential mechanism that satisfies $\epsilon$-DP releases query result $s$ calculated from data $D$ with probability  $\exp(u(s)\frac{\epsilon}{2\delta_u})\bigg/\bigintsss_{s'} \exp\!\left(u(s')\frac{\epsilon}{2\delta_u}\right) ds'$, where $\delta_u$ is the maximum change in score $u$ with one element change in data $D$.
\end{defn}

\subsection{STatistical Election to Partition Sequentially (STEPS)}\label{sec:steps}

\subsubsection{Overview}
STEPS aims at privately selecting a decomposition sequence  of the joint distribution  of $p$ attributes $\mathbf{X}=(X_1,X_2,\ldots,X_p)$ in the  data, and sanitizing each component in the decomposition  sequentially to achieve better original information preservation for the attributes that are more important to the  practical or statistical understanding of the population-level signals in the  data.  The outcome of STEPS is a private joint distribution $f(\mathbf{X})$, from which synthetic data can be generated and released. 

The decomposition of the joint distribution  can be a ``full'' decomposition such as $f(\mathbf{X})=f(X_1)f(X_2|X_1)\ldots, f(X_p|X_1,\ldots,X_{p-1})$, or ``partial'' decomposition such as $f(\mathbf{X})=\!f(X_1) \\ f(X_2|X_1) f(X_3,\ldots,X_{p-1}|X_1,X_2)$ or
$f(\mathbf{X})=\!f(X_1,\ldots,X_j|X_{j+1},\ldots,X_p)f(X_{j+1},\ldots,X_p)$, etc. For example, suppose a data set has three attributes $X_1,X_2,X_3$, and the  decomposition sequence chosen by STEPS is $f(X_3)f(X_1|X_3)f(X_2|X_1,X_3)$. STEPS sanitizes the three components $f(X_3), f(X_1|X_3)$, and  $f(X_2|X_1,X_3)$, independently, taking into account the necessary equality constraints along the way.  Since all three components contain information on $X_3$, STEPS will aggregate all the sanitized information on $X_3$ when generating the private joint distribution; similarly for $X_1$, the information on which is contained in two queries ($f(X_1|X_3)$ and  $f(X_2|X_1,X_3)$), and STEPS will aggregate both sources of sanitized information on $X_1$ when generating the private joint distribution. Suppose that the sanitization of each component is allocated the same amount of privacy budget, since $X_3$ is involved in all three components in the decomposition, it receives the most amount of privacy budget collectively and its original information is expected to be better preserved than the other two; so does $X_1$ compared to $X_2$.

\subsubsection{Procedure}\label{sec:procedure}
The STEPS procedure has three main steps: data partitioning and hierarchical tree construction; sanitization and correction; and synthesis and release. While STEPS can deal with both categorical and numerical attributes, the numerical ones will be first cut into histogram bins and  be treated as categorical attributes afterwards until the last step of synthetic data generation.  

\begin{figure}[!htb]
\vspace{-6pt}
\centerline{\includegraphics[width=6in]{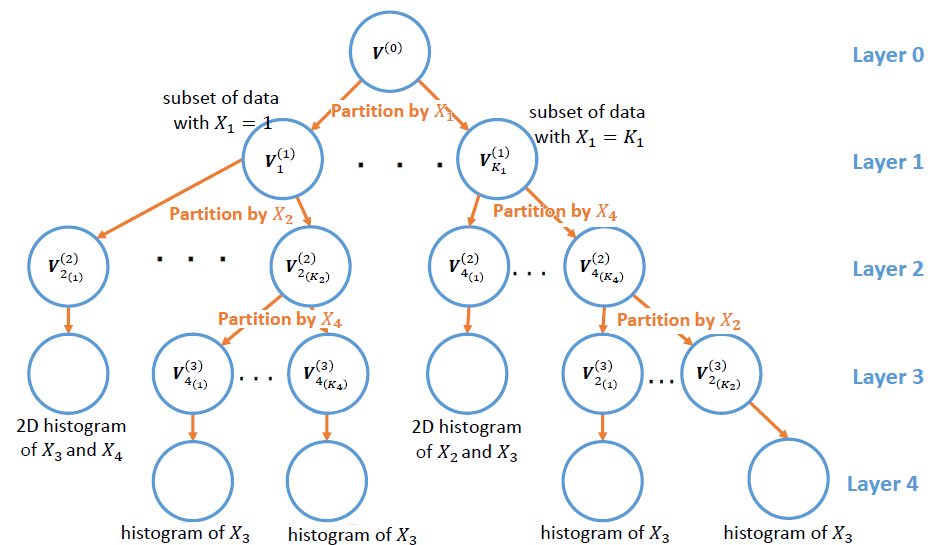}}
\caption{Illustration of an hierarchical tree built by STEPS in a 4-variable case.}\label{fig:steps}
\end{figure}

In the first step of data partitioning and hierarchical tree construction, STEPS selects a decomposition of the joint distribution $f(\mathbf{X})$. The decomposition sequence, once decided, can be represented in the format of a hierarchical tree $\mathcal{T}$. The closer an attribute is placed to the top of the tree, more statistics that involve that attribute will be sanitized the subsequent steps, the more privacy budget the attribute will receive collectively, and thus more original information is expected to be preserved for that attribute.   Figure \ref{fig:steps} shows an example of $\mathcal{T}$ built by STEPS when $p=4$. The layers are denoted by  $l=0,1,\ldots,L=4$, and the nodes in layer $l$ are denoted by $\vv^{(l)}$ and each of them contains the subset of data following the partitioning rule till that node along its branch of the tree. $\vv^{(0)}$, the node at the top of $\mathcal{T}$ contains the whole data set. $K_j$ represents the number of levels of variable $X_j$ for $j=1,\ldots,4$. Note that the order of the variables for partitioning may differ by branch, and  the branches do not have to be of the same length. The decomposition in the tree in Figure \ref{fig:steps} is $f(\mathbf{X})=\!f(X_3,X_4|X_2=1,X_1=1)f(X_2=1|X_1=1)f(X_1=1)+\ldots+
\sum_{j=1}^{K_4}f(X_3|X_2=K_2,X_1=1,X_4=j)f(X_4=j|X_2=K_2,X_1=1)f(X_2=K_2|X_1=1)f(X_1=1)+
\ldots+
f(X_2,X_3|X_1=K_1,X_4=1)f(X_4=1|X_1=K_1)f(X_1=K_1)+
\sum_{j=1}^{K_2}f(X_3|X_4=K_4,X_1=1,X_2=j)f(X_2=j|X_4=K_4,X_1=K_1)f(X_4=K_4|X_1=K_1)f(X_1=K_1)$.  

The sequence by which the data are partitioned or by which the attributes are aligned and placed on trees can be decided in at least two ways. First, one can leverage prior information or domain knowledge regarding the importance of the attributes, either practical or statistical, in a particular data set. The attributes  that are of more interest to practitioners or more important per the prior knowledge will be placed closer to the root (top) of the tree.  Applying external knowledge to aid certain aspects of developing a DP procedure (e.g., hyper-parameter tuning) has been used in other privacy-preserving approaches to help save on privacy cost and improve the utility of released results.  This approach does not cost any privacy of the current data as the information comes from outside. 
For the second approach, one may apply a metric that measures the  importance of each attribute to the statistical  understanding  the original data, such as  a quantity that quantifies the contribution of an attribute in explaining the variability of the data; attributes that explain more variability will be placed closer to the root of the tree. For example, information-theoretic model selection criteria such as AIC and BIC can be used to determine the  then the attribute selected to partition the data in a parent node is the one with the smallest AIC among all the univariate log-linear models fitted to the data. 
Since the metric is calculate from the original data and determines the tree structure, a portion of the total privacy budget should be allocated to this step to keep the determination of the partitioning sequence private.   

In the second step of sanitization and correction, a differentially private mechanism injects noises to the count of the data points contained in each of the nodes of the constructed tree $\mathcal{T}$. After the independent noise injection, the sum of the counts in the children nodes (denoted by $\mbox{succ}(v)$) are very likely not equal to the count in their parent node $v$, but they should. For example, a node $v$ contains all the  data points whose attribute ``race'' is ``white''. Say its original count of 100, and its children $\mbox{succ}(v)$ from further partitioning by ``gender''  include ``white female'' with an original count of 44  and ``white male'' of with an original count of 56. Suppose the sanitized counts are 120, 48, and 50, respectively, then ``white male'' and ``white female'' no longer add up to ``white''.  To ensure the equality constraints between $\sum \mbox{succ}(v)$ and $v$ in the hierarchical decomposition, a post-processing procedure similar to the UH approach is applied, with some necessary modifications. Specifically, $z$ is first calculated via Eqn (\ref{eqn:bottom}) as a weighted average of the directly sanitized count $n^*[v]$ and the indirectly sanitized count $\sum_{u\in \mbox{succ}(v)}z[u]$ summed from its children nodes in a bottom-up manner, and then the inconsistency is corrected to obtain the final consistent counts $\bar{n}^*$ via Eqn (\ref{eqn:top}).

\begin{align}
z[v^{(l)}]&=
\begin{cases}
n^*[v^{(l)}], & \text{if } l=L \text{ ($v$ is a leaf node)}   \\
\frac{b^{L-l}-b^{L-l-1}}{b^{L-l}-1}n^*[v^{(l)}]+ \frac{b^{L-l-1}-1}{b^{L-l}-1}\sum_{u\in \mbox{succ}(v^{(l)})}z[u] &  \text{if } l=L-1,\ldots,1
\end{cases};\label{eqn:bottom}\\
\bar{n}^*[v]&=
\begin{cases}
z[v^{(l)}],  & \text{if } l=L \text{ ($v$ is a leaf node)}   \\
z[v^{(l)}]+\frac{1}{b}(\bar{n}^*[w]-\sum_{u\in \mbox{succ}(w)}z[u]) &  \text{if } l=L-1,\ldots,1
\end{cases}.\label{eqn:top}
\end{align}

$w$ in  Eqn (\ref{eqn:top}) represents $v$'s parent; $b$ in Eqns (\ref{eqn:bottom}) and (\ref{eqn:top}) is the number of children per node, assumed to be the same across all nodes in $\mathcal{T}$ in the UH approach \citep{hay2010boosting}, and can be set by the data user ($b\ge2$); $b^{l}$ denotes $b$ to the $l$ power in Eqn (\ref{eqn:bottom}). A tree built by STEPS  might have branches of different length (e.g. Figure \ref{fig:steps}). To accommodate the requirement of the same $b$ across all nodes so that Eqns (\ref{eqn:bottom}) and (\ref{eqn:top}) still apply,  \emph{phantom} categories can be generated to make each attribute have the same number of categories as the attribute with the most categories. The ``original'' counts in the phantom categories are 0, and will remain at 0 during the sanitization and correction process.

In the last step of synthesis and release, if all attributes are categorical, we can release directly the sanitized counts of the nodes in the bottom layer of the tree; otherwise, we can apply uniform sampling to draw synthetic values for the numerical attributes in each bin of the formed histograms in each node.

\subsubsection{Algorithm}
The algorithmic steps of the STEPS procedure are given in Algorithm \ref{alg:steps}. 
\begin{algorithm}[!tbh]
\caption{STatistical Election to Partition Sequentially with $L$ layers (STEPS-$L$)}\label{alg:steps}
\begin{algorithmic}[1]
\State \textbf{Input}: number of layers $L\; (\le p)$ (layer 0 is not counted toward $L$ which contains all data points); overall privacy budget $\epsilon$; the portion $r\epsilon$ for $r\in[0,1)$ allocated for determining decomposition; utility function $u$ (e.g., AIC) for the exponential mechanism and its sensitivity $\delta_u$; number of synthetic sets $m$; original data $D$.
\State \textbf{Output:} $m$ synthetic sets $\tilde{D}^{(1)},\ldots,\tilde{D}^{(m)}$
\State \textbf{If} there is a predefined partitioning sequence, $r=0$ and build tree $\mathcal{T}$ according to the predefined sequence.
\State \textbf{Else} define $\mathcal{A}^{(0)}_{v^{(0)}[1]}$ that includes all $p$ attributes in $D$.
\State \hspace{10pt} \textbf{For} $l=0$ to $L$
\State \hspace{20pt} \textbf{For} $k=1$ to $K^{(l)}$  ($K^{(l)}$ is number of nodes in layer $l$)
\State\hspace{26pt}$\bullet$ Apply a univariate loglinear model to the data contained in node $\vv^{(l)}[k]$ on each\\ \hspace{33pt} attribute from its availability set $\mathcal{A}^{(l)}_{\vv^{(l)}[k]}$, and calculate $u$.
\State\hspace{26pt}$\bullet$ Choose a partitioning variable via the exponential mechanism  with privacy  budget \\ \hspace{33pt}  $r\epsilon/(mL)$. Denote the selected attribute by $X_{j}[k]$.
\State \hspace{26 pt}$\bullet$ Exclude $X_{j}[k]$ from the availability sets for the all children nodes of $\vv^{(l)}[k]$ in layer \\ \hspace{33 pt}   $l+1$.
\State \hspace{20 pt} \textbf{End For}
\State \hspace{10 pt} Obtain the full histogram over all the attributes in $\mathcal{A}^{(L)}_{\vv^{(L)}}$ for all nodes in layer $L$.
\State \hspace{10 pt} \textbf{End For}
\State \textbf{End Else}
\State  Sanitize all nodes in the generated tree $\mathcal{T}$ (e.g. via the Laplace mechanism), and apply the inconsistency correction in Eqns (\ref{eqn:bottom}) and (\ref{eqn:top}) to obtain the final sanitized counts $\bar{\mathbf{n}}^*$. 
\State  If all attributes are categorical, release the sanitized counts of the nodes in the bottom layer of $\mathcal{T}$; otherwise, apply  uniform sampling to draw synthetic values for the numerical attributes on which histograms are formed.
\end{algorithmic}
\end{algorithm}

The input to the STEPS algorithm includes a user-specified number for the partition layers $L$. Though the case of $L>p$ would be possible by allowing the levels of an attribute to span across multiple layers, this does not seem necessary from a data synthesis perspective, especially when the partitioning sequence is determined statistically. Therefore, we require that a variable used in partitioning data in earlier layers is no longer available along that branch; in other words, the set of available attributes $\mathcal{A}^{(l)}_{\vv^{(l)}[k]}$ for partitioning the $k$-th node  $\vv^{(l)}[k]$ in layer $l$ will shrink as the tree grows. When $L<p$, there are more than one attributes in $\mathcal{A}^{(L)}_{v^{(L)}[k]}$ for the $k$-th nodes in layer $L$, but no more partitioning will be applied and the leaf nodes will be made of the cells from the full cross-tabulation over all the remaining attributes in $\mathcal{A}^{(L)}_{\vv^{(L)}[k]}$.

Regarding the choice of $L$, a larger $L$ is associated with more computational complexity and less privacy budget per layer. On the other hand, since the counts in then top layers are weighted averages per Eqn. (\ref{eqn:top}), the decrease in the privacy budget per layer could be offset or even trumped by the information gain aggregated over multiple sources of information. Given the above reasoning, we conjecture that there exists an optimal $L$ based on the trade-off between privacy budget and data utility. Users may try different values of $L$ and pick a $L$ leads to the best utility among all available choices; but the procedure of choosing $L$ itself, if using the original data, costs privacy. For future work, we will look into devising a stopping rule for tree building rather than choosing $L$ beforehand. 

If the tree structure is suggested by external knowledge, then the tree building costs no privacy; if the data set itself is used to suggest the structure of a tree, the overall privacy budget will be split between building and sanitizing the hierarchical tree. Specifically, a certain portion $r$ of the overall privacy, which is further split into $L$ layers, will be allocated to find the optimal split variable for a node in each layer via the Exponential mechanism; the rest of the privacy budget ($1-r$) can be used to sanitize the node counts in the constructed tree, which is also further split into $L$ partition layers. The optimal $r$, a hyperparameter, likely depends on the data; but users could preset a $r$ value they are willing to spend on tree building if they do not want to spend budget to pick $r$. Since different nodes in the same layer do not have overlapping information, choosing the split variables and sanitation of the nodes follow the parallel composition principle. Taken together, with $m$ synthetic data sets, and if each layer receives the same amount of budget for count sanitization, then each node count in layer $l$ receives a budget of $(1-r)\epsilon/(mL)$.

When the exponential mechanism is used to privately choose a partitioning variable from the availability set  $\mathcal{A}^{(l)}_{\vv^{(l)}[k]}$  for node $\vv^{(l)}[k]$, it samples attribute  $j\in\mathcal{A}^{(l)}_{\vv^{(l)}[k]}$ with probability \vspace{-3pt}
\begin{equation}\label{eqn:exp} \vspace{-3pt}
\frac{\exp\left(u_j(\vv^{(l)}[k])\epsilon^{(l)}_{k}/(2\delta_u) \right)}{\sum_{j'\in\mathcal{A}^{(l)}_{\vv^{(l)}[k]}} \exp\left(u_{j'}(\vv^{(l)}[k]) \epsilon/(2\delta_u)\right)},    \vspace{-3pt} 
\end{equation}
where $\delta_u$ is the maximum change in the utility function $u$ with one element change in the data contained in $\vv^{(l)}[k]$, and $\epsilon^{(l)}_{k}$ is the privacy budget node $\vv^{(l)}[k]$ receives for employing the exponential mechanism. Users may choose any reasonable utility function $u$ in Algorithm \ref{alg:steps}, but an easy and useful choice is AIC, whose  $\delta_u=2$ per Lemma  \ref{lem:AIC}.
\begin{lem}\label{lem:AIC} the sensitivity $\delta_u$ is 2 when the utility function $u$ in the exponential mechanism is AIC from a  univariate log-linear model with a single attribute.
\end{lem}
\begin{proof}
AIC $=-2\log(\mathcal{L}) + 2K$, where $\mathcal{L}$ is the likelihood function and $K$ is the number of the parameter of the model. For the univariate log-linear model with an independent variable of $K$ levels, $\mathcal{L}=\frac{n_v!}{\prod_{k=1}^K n_{v,k}! } \prod_{k=1}^Kp_k^{n_{v,k}}$ and $\log(\mathcal{L})=\log(n_v!)-\sum_{k=1}^K \log(n_{v,k}!)+\sum_{k=1}^K n_{v,k}\log(p_k)$, where $n_v$ is the data size, and $n_{k,v}$ is the count in level $k$ of that attribute.  Removing one element from $v$ leads to a loss of one observation in one of the $K$ categories, say level $j$. The likelihood function based on the $n_v-1$ observation is 
$L'=\frac{(n_v-1)!}{(n_{v,j}-1)!\prod_{k\ne j} n_{v,k}!} p_j^{n_{v,j}-1} \prod_{k\ne j} p_k^{n_{v,k}} $ and  $\log(\mathcal{L}')=\log((n_v-1)!)-\log((n_{v,j}-1)!)-\sum_{k\ne j} \log(n_{v,k}!)+n_k\log(p_k)+\sum_{k=1}^K n_{v,k}\log(p_k)$.  Therefore, the change in AIC is $\Delta$AIC$=-2\log(\mathcal{L})+2K+2\log(\mathcal{L}')-2K'$ with the removal of one attribute. Plugging the log-likelihoods, we have $\Delta$AIC$=\log(n_v)-\log(n_{v,j})+n_{n,j}\log(p_j)-(n_{n,j}-1)\log(p_j) +2K-2K'$. The true parameter $p_j$ is unknown and its maximum likelihood estimate is $n_{v,j}/n_v$ and  $(n_{v,j}-1)/(n_v-1)$ before and after removing an observation. Therefore, $\Delta$AIC 
$=\log(n_v)-\log(n_{v,j})+n_{v,j}\log(n_{v,j}/n_v)-(n_{n,j}-1)\log((n_{v,j}-1)/(n_v-1)) +2K-2K'= (n_{v,j}-1)\log(n_{v,j}(n_v-1)/(n_v(n_{v,j}-1))+2K-2K'$. When $n_j\ge2$, $K=K'$ and thus $\Delta$AIC 
$=(n_{v,j}-1)\log(n_{v,j}(n_v-1)/(n_v(n_{v,j}-1)) \in(0,1)$. When $n_j=1$, then $K'=K-1$, and $\Delta$AIC$=2$. Taken together, $\delta_u$ with AIC is 2.
\end{proof}

Plugging $\delta_u =2 $ in Eqn (\ref{eqn:exp}), then attribute $j$ is sampled with probability
\begin{align}
&\textstyle \exp(-\mbox{AIC}_j\epsilon^{(l)}_{k}/4)/\sum_{j'}\exp\left(-\mbox{AIC}_{j'}\epsilon^{(l)}_{k}/4\right)\label{eqn:AIC2}
\end{align}
If an AIC has a large negative value, direct application of Eqn (\ref{eqn:AIC2}) may lead to overflow problems in computation. It is thus suggested to replace AIC$_j$ and AIC$_{j'}$ above by their differences from  $\max_{j'}\mbox{AIC}_{j'}$ in practical implementation.

Algorithm \ref{alg:steps} may have other variants. One such variant is that the data  in a node can be partitioned by a group of variables rather than by a single variable, if that group of variables are similar in their importance measures. For example, suppose there are 5 variables in a data set, each associated with their respective importance measure. Suppose $L$ is pre-set at 2.  There are 15 ways of splitting the 5 variables into two clusters if each cluster has to have at least one variable. To choose which splitting scheme to use, we may use the $l_1$ difference between the two clusters  on the average importance measures across the variables in each cluster as the utility function for the exponential mechanism. The clustering scheme with the largest $l_1$ difference has a higher probability being chosen. Once the exponential mechanism chooses a clustering, the group with the higher average importance measure will be put in layer 1 and the other will be in layer 2. This generalized splitting scheme is what is used in the simulation studies in Sec \ref{sec:simulation}. 

Algorithm \ref{alg:steps} suggests releasing multiple synthetic sets ($m>1$) so to capture the sanitization and synthesis uncertainty to yield valid statistical inferences based on the released synthetic data. While releasing multiple sets is not the only way to capture the uncertainty, it is likely the most straightforward and simplistic way in terms of practical implementation. Statistical inference for analysis over the multiple sets can be obtained via the formulas given in \citet{liu2016model} and \citet{bowen2020comparative}. To preserve the overall DP in the release of $m$ sets, each synthetic set is allocated $1/m$ of the total privacy budget $\epsilon$ per the sequential composition. In terms of the choice for $m$, $m$ too large  will lead too much information loss for a single synthetic set to be remedied by aggregating over $m$ sets of information; and $m$ too small might not adequately propagate the inherent uncertainty from sanitization and synthesis. Our empirical studies suggest $m=3$ to $6$ is a good choice \citep{liu2016model}.

\subsection{STEPS vs UH and PrivBayes}\label{sec:UH}
The STEPS procedure focuses on constructing a differentially private joint distribution, in relatively high dimensional settings, from which synthetic data are generated. It aims at  preserving more information for important variables, where ``important'' can be defined either statistically or per domain knowledge.  By contrast, most existing partitioning approaches focus on improving the accuracy of marginal counts in low-dimensional setting. In addition, they often allow the same attribute to span multiple layers whereas different layers in STEPS contain non-overlapping attributes.  

Among the existing partitioning-based DP approaches, STEPS relates to the UH the most as it uses the inconsistency correction rules developed in the latter. Exploring and maintaining the inherent consistency constraints in the UH procedure helps improve the accuracy of low-dimensional marginal queries. The UH is mostly studied in the context of a single attribute, where the high-level nodes present large-range queries and the leaf nodes in the lowest level are the finest categories/bins for the attribute. Though it can be applied to multidimensional data, its benefit over the one-step Laplace sanitizer seems to diminish over increasing dimensionality \citep{qardaji2013understanding,qardaji2013geo}. In addition, the UH approach has an exponential time complexity $O(b^L)$ in $L$ for a given $b$ (Eqns (\ref{eqn:bottom}) and (\ref{eqn:top})), implying a large $L$ would dramatically increase the computational time \citep{qardaji2013understanding}.

Another related DIPS method to STEPS is PrivBayes \citep{PrivBayes}, which also relies on the decomposition of the full distribution  $f(\mathbf{X})$ of the attributes in the data. However, PrivBayes is different from STEPS in several aspects. First, PrivBayes uses a low-dimensional distribution to approximate the full distribution  $f(\mathbf{X})$ by exploring the conditional independence among the attributes. In other words, PrivBayes is model-based approach (though Bayesian networks) to represent the signals in the data. As such, the sanitized data are subject to bias if the Bayesian network does not provide a good fit to the data. In contrast, STEPS does not imposes a model on the data and it always sanitizes the full-dimensional empirical  distribution, though it may leverage modelling to determine which layer to put a certain attribute and emphasizes the preservation of more information on more important attributes.  Second, the PrivBayes injects noises to all parent-children node pairs in parallel to derive the differentially private approximate probability mass function to $f(\mathbf{X})$ from which synthetic data can be generated; correction for inconsistency is not needed during pr after the sanitation.    

\subsection{A General Utility Metric}\label{sec:specks}
To assess the similarity between synthetic data and actual data, we develop the SPECKS (\textbf{S}ynthetic data generation; \textbf{P}ropensity score matching; \textbf{E}mpirical \textbf{C}omparison via the \textbf{K}olmogorov-\textbf{S}mirnov distance) metric. SPECKS is a propensity-score-based general utility measure and can be used to compare the similarity of two data sets of the same structure of any dimension without making assumptions on the distributions of the attributes. The procedure of SPECKS is given below. Each step is straightforward and easy to implement.

\begin{itemize}
\item[1) ] Combine the original and synthetic data, each of size $n$. Create an indicator variable $T$ where $T_i=1$ if  record $i$ is from the synthetic data and $T_i=0$ otherwise  for $i=1,\ldots, 2n$.
\item[2) ] Calculate the propensity score for each record $i$, $e_i=\Pr(T_i=1|\x_i)$, through a classification algorithm, with the data attributes as input features. 
\item[3) ] Calculate the empirical CDFs of the propensity score, $\hat{F}(e)$ and $\tilde{F}(e)$, for the actual and the synthetic groups, separately.
\item[4) ] Compute the Kolmogorov-Smirnov (KS) distance $d=\sup_{e}|\tilde{F}(e)-\hat{F}(e)|$ between the two empirical CDFs (if multiple synthetic data sets are generated, the average KS distance over the multiple sets is taken).
\end{itemize}

If the synthetic data preserve the original information well, then the observations from the two groups are indistinguishable and a small KS distance between the original and synthetic empirical CDFs is expected.   In the second step of the SPECKS procedure, any classifier (e.g., logistic regression, random forests, SVM) can be used. In the case of logistic regression, the model covariates may include the main effects of the data attributes or interaction terms among the attributes. This implies that the propensity scores will vary by classifier. \citet{bowen2019comparative} note that different classifiers measure different types of distributional similarity. For instance, a logistic regression model with only main effects measures data similarity in the marginal distributions (simultaneously) whereas a more complex CART model can measure similarity in high-order dependency among the attributes.

Compared to other propensity-score-based utility measures or discriminate-based methods, SPECKS has similar steps such as the calculation of propensity scores, but differs in how it formulates the final utility metric from the estimated propensity scores (Steps 3 and 4 above). In \citet{sakshaug2010synthetic}, the propensity scores are discretized based on how the Chi-squared test is formulated. \citet{woo2009global} calculate the mean squared error (MSE) of the calculated propensity score vs the true proportion of synthetic cases. \cite{snoke2018general} normalize the MSE statistic by its expected null value and standard deviation, which helps with its interpretability and also seems to be more sensitive in telling the synthetic data apart from the actual data. However, the derivation of the expected null  value and standard deviation is driven large-sample assumptions. In contrast, SPECKS utilizes the KS distance which is the maximum distance of two empirical CDFs, and thus considers the worst-case separation between the synthetic data and the actual data.

\section{Simulation Studies}\label{sec:simulation}
In this section, we implement the STEPS method in simulated data to generate differentially private synthetic data, and compare the statistical utility of the synthesized data between STEPS vs. a modified version of the UH and PrivBayes. 

\subsection{Choice of Methods for Comparison}
The reasons for choosing the UH and PrivBayes to compare with STEPS are given below. The same justifications apply to the case study in Sec \ref{sec:voter}.

First, the UH and PrivBayes methods are the most related methods to STEPS as presented in Sec \ref{sec:UH}. STEPS uses the inconsistency correction rule of the UH  In addition, \citet{qardaji2013understanding} suggest that the UH  generally is the best data-independent algorithm and achieves lower error for range queries from one-dimensional histograms, compared to many data-independent methods. When implementing the procedure, we made some modification to the original UH to accommodate computational constraints and to handle multi-dimensional data. Both STEPS and PrivBayes build a differentially private joint distribution among the data attributes  from which synthetic data are generated. Similar to STEPS, PrivBayes deals with numerical attributes by discretizing them into histogram bins. 

Though Privelet and DPcube, originally  developed for answering range queries from one- or two- dimensional histograms, can be potentially used for data synthesis, we do not examine them in the simulation studies as neither offers better performance than the UH approach in general even in the low-dimensional setting \citep{hay2016principled}. MWEM \citep{hardt2012simple} aims to obtain differentially private queries rather than generating synthetic data from an empirical distribution. To use MWEM for data synthesis, choosing a good set of queries that represent the information of the original data is vital. Even with a good set of queries to start with, it does not necessarily outperform the flat Laplace sanitizer when generating synthetic data \citep{cipher}. In addition, MWEM is proved to be inconsistent and its performance highly depends on the number of iterations \citep{hay2016principled, dawa, cipher}.   Though the DAWA method \citep{dawa} beats other methods including DPcube, MWEM, and Privelet for low-dimensional range queries, the good performance replies on allocating some privacy budget to identify an optimal bucketing scheme on the numerical attributes before the sanitization. Given most of the attributes in our simulation and case studies are either categorical or ordinal with limited number of levels, the advantages of DAWA will not be brought into full strength. Though the model-dependent MODIPS method can deal with all types of data in theory, it has a couple of limitations for practical implementation, such as that not every sufficient statistic is easy to sanitize and a mis-specified model would generate biased synthetic data. DPcopula uses copula to model pairwise dependency among attributes, based on which data are synthesized; the method assumes continuous marginal CDF; and the synthetic data can also be sensitive to the type of copula employed.   

\subsection{Simulation Setting}
We simulated data from two linear regression models. For each model, we simulated 200 data sets. Model 1 is main-effect model: $Y=\beta_0+\sum_{j=1}^{10}\beta_jX_j+\epsilon$ with $\epsilon\sim N(0,1)$. $X_1$ to $X_4$ are categorical predictors with 2, 3, 4 and 5 levels respectively, and the true $\bs{\beta}$ values are $0,0.5,(0.5,-0.8),(1,0.5,0.5),(0.5,0.7,0.8,0.5)$, respectively. Model 2 is a saturated model: $Y=\bs\beta^T \mathbf{X}_{1234}+\epsilon$  with $\epsilon\sim N(0,1)$. $\mathbf{X}_{1234}$ refers the 4-way interaction term among $X_1, X_2,X_3,X_4$, and $\bs\beta$ is 120-dimensional with each of its elements  sampled from a standard normal distribution and then fixed for all 200 repeats. The two models represent the two extremes of a wide spectrum of possible models with 4 predictors.

We examined two sample sizes scenarios, $n=10,000$ and $n=4,000$, in each model. We set privacy budget $\epsilon$ at $0.5, 2, 5, 10$, respectively. For each of the 12 simulation scenarios (2 models at 2 sample sizes with 3 $\epsilon$ values), four sets of synthetic data were generated  by each of the three DIPS approaches (STEPS, UH, and PrivBayes). The numerical attribute $Y$ was cut into 15 bins before the application of each method, and was treated as categorical until the last step before data release (refer to line 18 of Algorithm \ref{alg:steps}).

For STEPS, we split the total $\epsilon$ in a $1:9$ ratio between building the tree and sanitizing node counts for the tree. We used AIC as the utility function in the exponential mechanism for choosing partitioning attributes. Since the group of variables $(X_1,X_2,X_4)$ had much smaller AIC than the group $(X_3,Y)$, we used the former group as the partitioning variables in Layer 1, and put the second group $(X_3,Y)$ in Layer 2 to form a tree with $L=2$. The Laplace mechanism was applied to sanitize the node counts in the tree. In addition, We created phantom categories for $X_1$ to $X_4$ to match the number of bins $b$ in discretized $Y$ so that the inconsistency correction formulas in Eqns (\ref{eqn:bottom}) and (\ref{eqn:top}) can be applied  (see Sec \ref{sec:procedure}). For the UH, we modified the original approach to allow unequal number of children per node. In fact, the modified UH is more of a STEPS procedure with a random partitioning sequence and $L=4$. For PrivBayes, we applied the Python codes by \citet{ DataSynthesizerpaper}, available on GitHub \citep{DataSynthesizercodes}. The degree of the Bayesian network was set at 2 in all the simulation scenarios. The total $\epsilon$ was split in half between choosing a Bayesian network vs. sanitizing the joint distribution among the attributes (the default setting).

We run the true underlying linear regression models on each synthetic set and the inferences on the regression coefficients were summarized via the combination rule given in \citet{liu2016model} and \citet{bowen2020comparative}. For Model 2, the ridge regression was applied given the large number of parameters. The bias, root mean squared error (RMSE), and coverage probability (CP) of the 95\% confidence interval (CI) were obtained for each regression coefficient in each model. We also applied the estimated private models based on the synthetic data to predict $Y$ in an independent testing data set of $n=100$, and reported the prediction MSE. 

\subsection{Results}
\begin{figure}[!htb]
\small{para. est. bias$\;\quad$}
\raisebox{-0.5\height}{\includegraphics[scale=0.552]{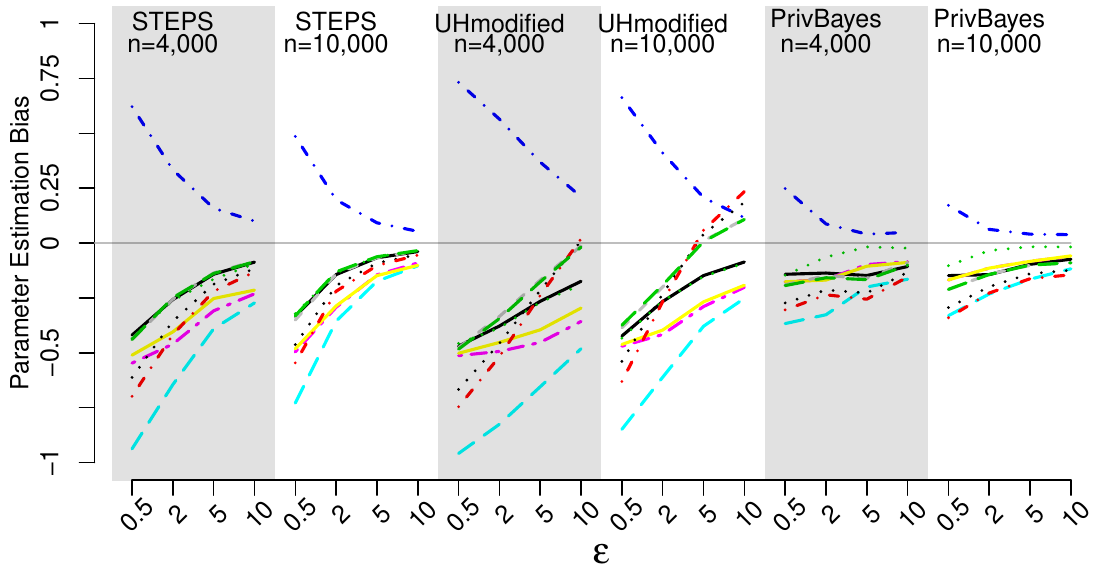}}\\
\small{para. est. RMSE}
\raisebox{-0.5\height}{\includegraphics[scale=0.552]{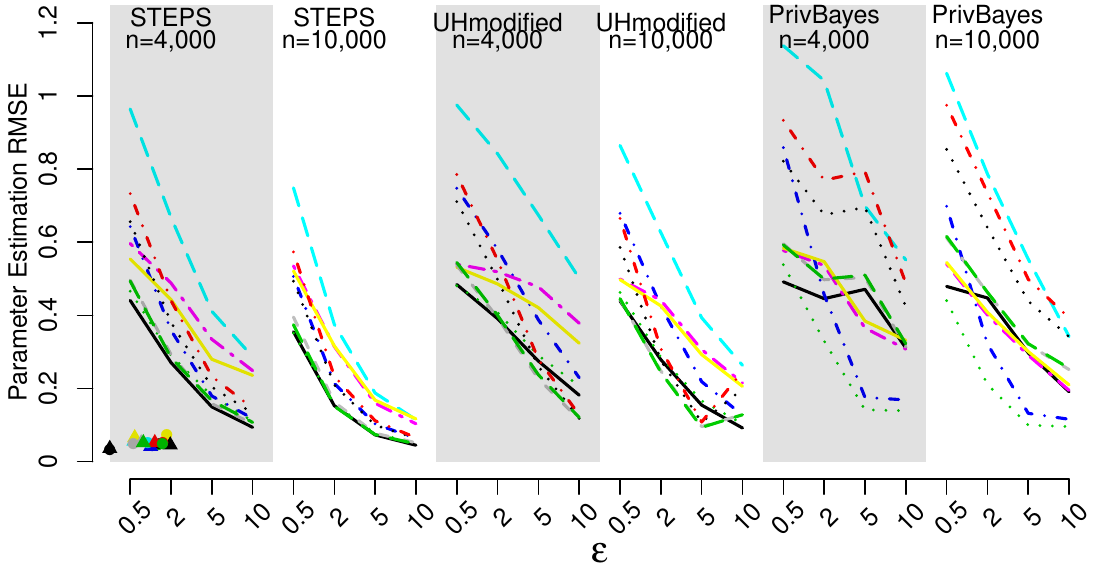}}\\
\small{CP of 95\% CI$\;\quad$}
\raisebox{-0.5\height}{\includegraphics[scale=0.552]{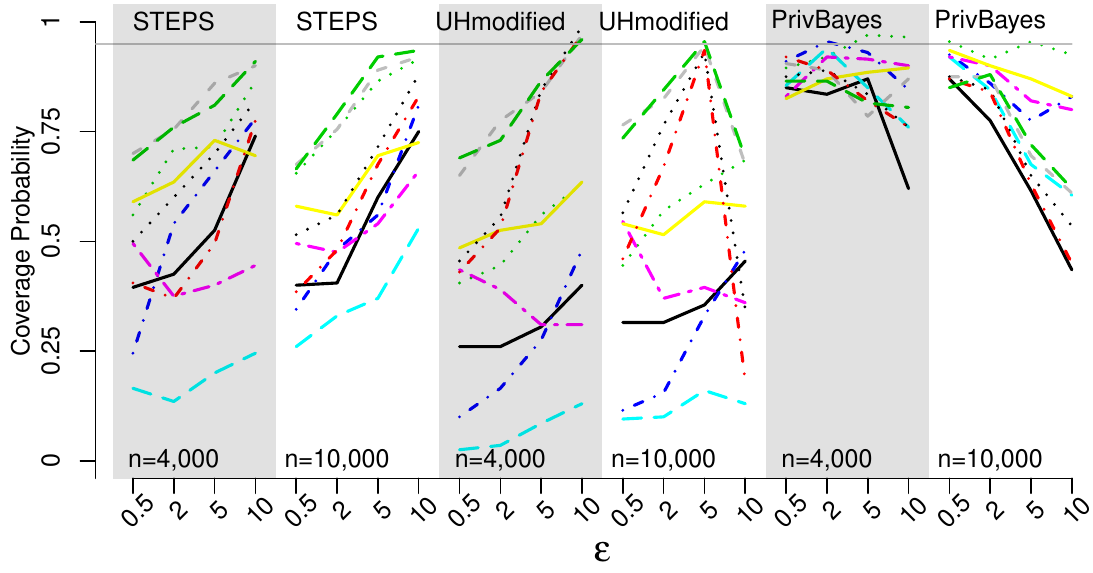}}\\
\small{prediction RMSE}
\raisebox{-0.5\height}{\includegraphics[scale=0.552]{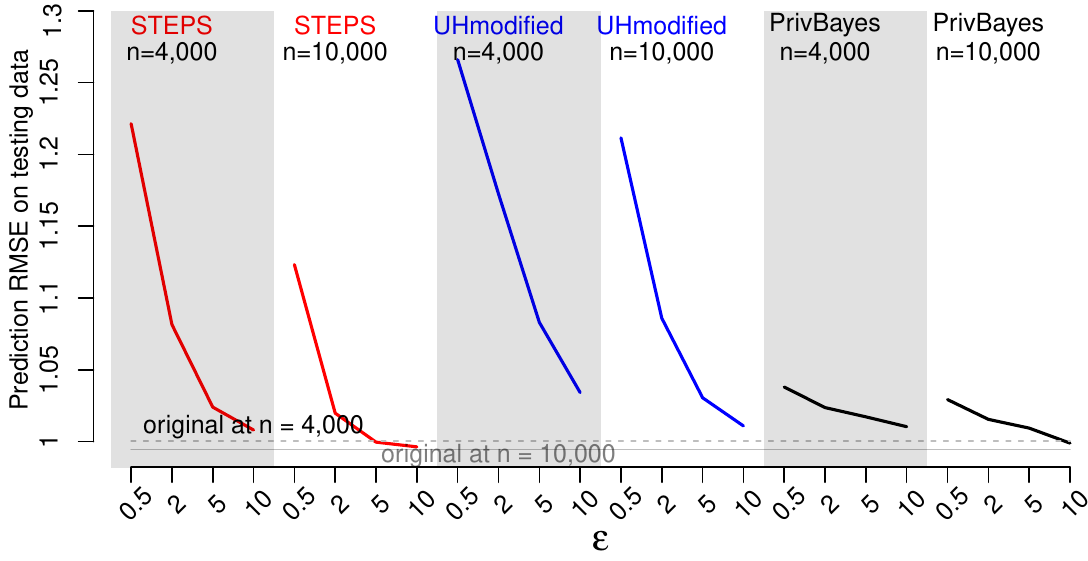}}\vspace{-9pt}
\caption{Inferences based on differentially private synthetic data in the simulation study for Model 1. Each line represents a different regression coefficient in Model 1 in each of the top 3 plots.} \label{fig:maineffect}\vspace{-12pt}
\end{figure}

\begin{figure}[!htb]
\small{para. est. bias$\;\quad$}
\raisebox{-0.5\height}{\includegraphics[scale=0.53]{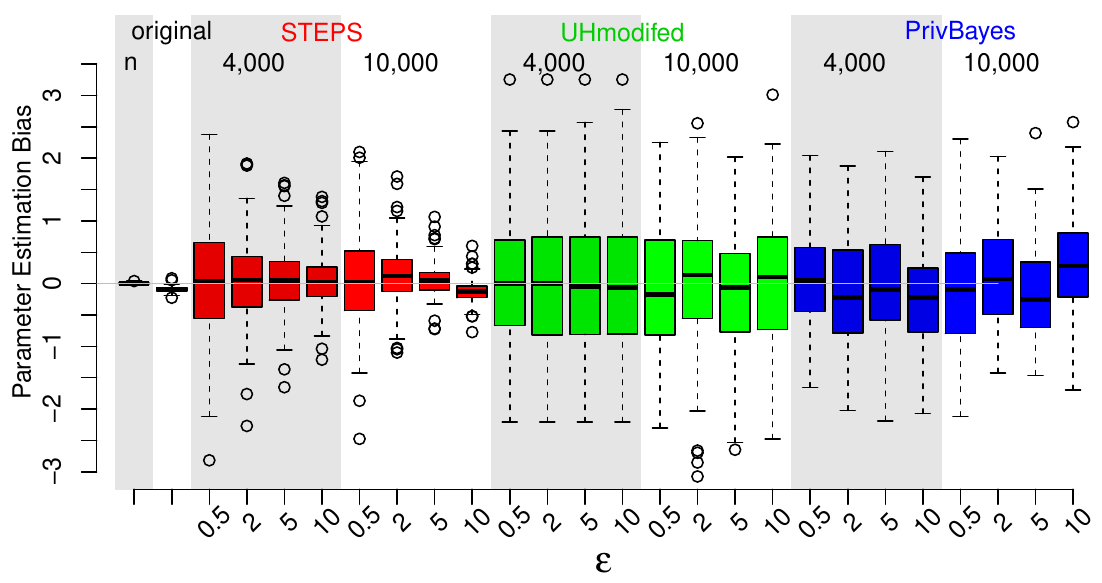}}\\
\small{para. est. RMSE}
\raisebox{-0.5\height}{\includegraphics[scale=0.53]{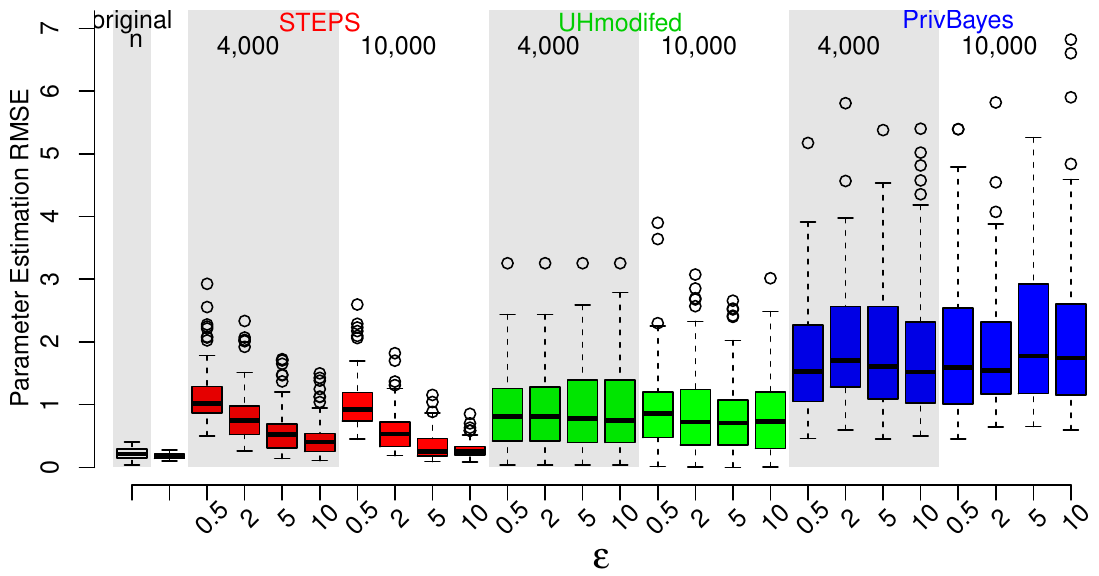}}\\
\small{CP of 95\% CI$\;\quad$}
\raisebox{-0.5\height}{\includegraphics[scale=0.53]{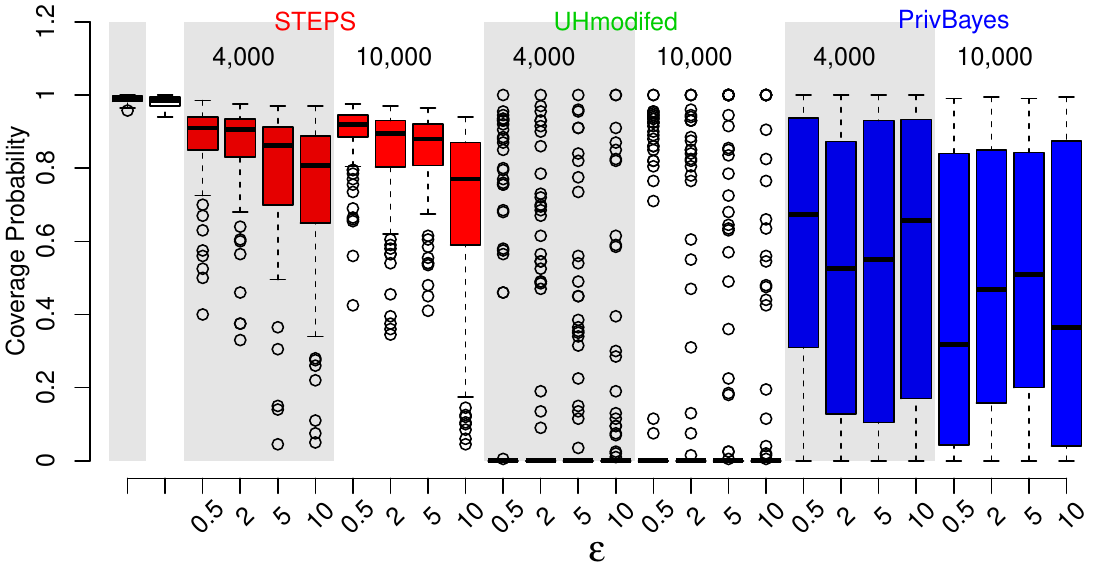}}\\
\small{prediction RMSE}
\raisebox{-0.5\height}{\includegraphics[scale=0.53]{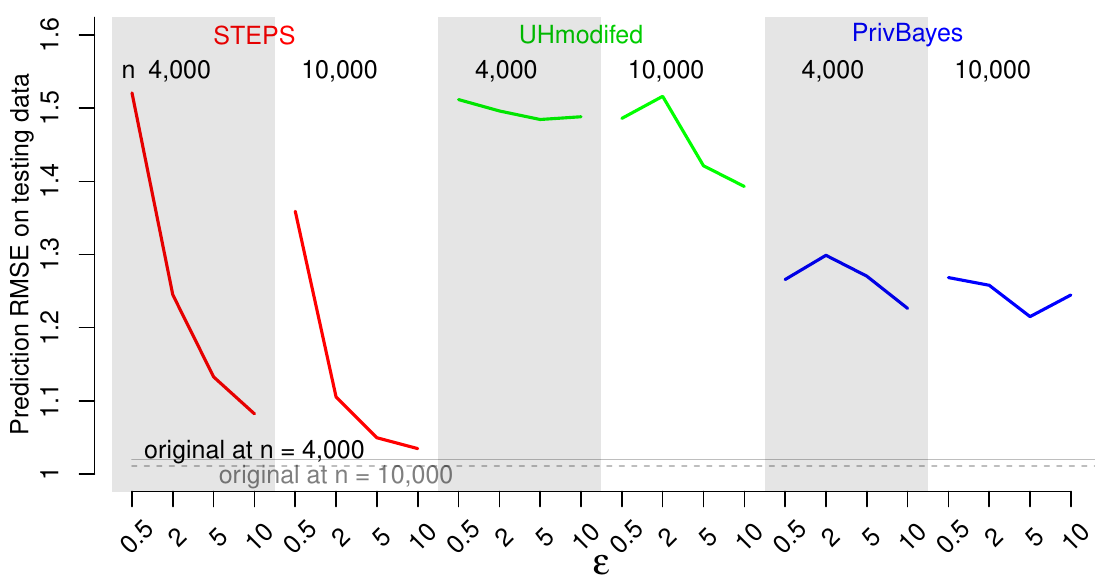}}\vspace{-9pt}
\caption{Inferences based on differentially private synthetic data in the simulation study for Model 2. In the  bias figure, each box plot represents the distribution of the biases over the 120 model parameters in a specific simulation scenario; similarly for the RMSE and CI.} \label{fig:saturated}\vspace{-12pt}
\end{figure}

The results are presented in Figures \ref{fig:maineffect} and \ref{fig:saturated} for the main-effect model  (Model 1) and the saturated model (Model 2), respectively. 

When the true model contains only the main effects, PrivBayes performs the best in parameter estimation bias, CP, and  prediction RMSE and the worst in parameter estimation RMSE; STEPS performs the best in  parameter estimation RMSE; the modified UH is the worst in parameter estimation bias, CP, and the prediction RMSE. As expected, the inferences improve as $n$ or $\epsilon$ increases;  the only exception is the trend of CP across $\epsilon$ for PrivBayes at both $n$ scenarios, where  more deviation from the nominal 95\% level is observed as $\epsilon$ gets larger. This counter-intuitive observation is likely due to the $Y$ discretization. Specifically, when $\epsilon$ is large, the main source of information loss comes from the (deterministic) discretization in $Y$ and the main source of variation across the multiple synthetic sets is the uniform sampling from each sanitized bin, rather than the noises introduced by sanitization. When $\epsilon$ is small, the main source of information loss and variability across multiple synthetic sets is  sanitization, and releasing multiple synthetic sets is designed to take into account that source of variation and the CP is rather better than that at large $\epsilon$.

When the true model is saturated, STEPS performs the best overall; PrivBayes is the worst in parameter estimation RMSE, and the modified UH is the worst in CP and prediction RMSE. For STEPS, the inferences improve as $n$ or $\epsilon$ increases as expected, with the only exception in the case of CP across $\epsilon$ at both $n$ scenarios -- likely due to the same reason as discussed above in the case of PrivBayes in the main-effect model. For the modified UH, the inferences get better as $n$ increases, but the trend is not obvious across $\epsilon$. For PrivBayes, the trends across $n$ and $\epsilon$ are not obvious in any of the examined metrics.  


\section{Application to the 2002-2012 CPS Youth Voter Data}\label{sec:voter}

\subsection{Data Description}
The 2000-2012 Current Population Survey (CPS) is downloaded from the Harvard Dataverse. The CPS is the primary source of labor force statistics for the United States population. In election years, the CPS also collects data on reported voting and registration, providing a large and nationally representative sample with coverage of both registered and non-registered individuals, and reports statistics on the voter turnout, age, race, among others. The voting and registration data from the CPS are frequently used and cited by various news outlets such as Time \citep{Time}, New York Times \citep{NYT}, Fortune \citep{Fortune}, and Newswise \citep{Newwise}, among others. 

In this application, we focus on the youth voter subset. The data was used to examine the effect of preregistration laws on turnout among young voters \citep{Holbein2016}. The data set has $n=44,821$ observations and $p=15$ variables  (Table \ref{tab:datlist}), with sensitive attributes such as Family Income and Voted and pseudo-identifiers such as gender, race information, and location information that can be used by adversaries to identify subjects or to link to other databases. Therefore, it is important to ensure that the private information is protected before releasing the data.

\begin{table}[!htb]
\centering
\resizebox{0.9\textwidth}{!}{
\begin{tabular}{@{}L{2.1in}@{}| L{4.1in}@{} }
\hline
Variable & category (percentage)\\
\hline
Voted &yes (34.7), no (63.3) \\
Preregistration State & yes (10.5), no (89.5)\\
Age (years) & 18 (20.5), 19 (19.7), 20 (20.0), 21 (20.0), 22 (19.8)\\
Married &  yes (7.5), no (92.5)\\
Female & yes (50.7), no (40.3)\\
Family Income$^\ddag$ & 14 levels (3.1, 5.9, 26.5)$^\ddag$ \\
College Degree & yes (3.1), no (96.9)\\
White &yes (69.3), no (31.7)\\
Hispanic &yes (14.6), no (85.4)\\
Registration Status & yes (52.4),  yes (47.8)\\
Metropolitan Area &yes (77.9), no (22.1)\\
Length of Residence & $<$1 month (2.7),
1-6 months (20.0),
7-11 months (6.9),
1-2 years (15.0),
3-4 years (9.8),
$\ge5$ years (45.6)\\
Business/Farm Employment & yes (12.6), no (12.62)\\
In-Person Interview &yes (36.7), no (63.3)\\
DMV Registration &yes (12.7), no (87.3)\\
\hline
\end{tabular}}
\resizebox{0.9\textwidth}{!}{\begin{tabular}{l}
\footnotesize For Family Income, the minimum,  medium and maximum percentages among the 14 levels are listed.\\
\hline
\end{tabular}}
\caption{List of variables and their values from youth voter subset in the 2000-2012 CPS data.}
\label{tab:datlist}
\end{table}


\subsection{Implementation}\label{sec:stepsinvoter}
Similar to the simulation studies, we compare the STEPS procedure in data utility against the modified UH approach and PrivBayes. \citet{hay2016principled} observe that when $n$ or $\epsilon$ is large, it is unlikely that any of the more complex algorithms will beat a simpler and easier-to-deploy flat algorithm. Similar observations regarding the Laplace sanitizer are also obtained in \citet{bowen2020comparative}, especially when $p$ is large. Given that the voter data set has a large $n$ ($44,821$) and a relatively large $p$ (15), it will be of interest to see how STEPS fares against the simple flat Laplace sanitizer, which is  included as another benchmark.   

\begin{figure}[!htb]
\centerline{\includegraphics[width=6in]{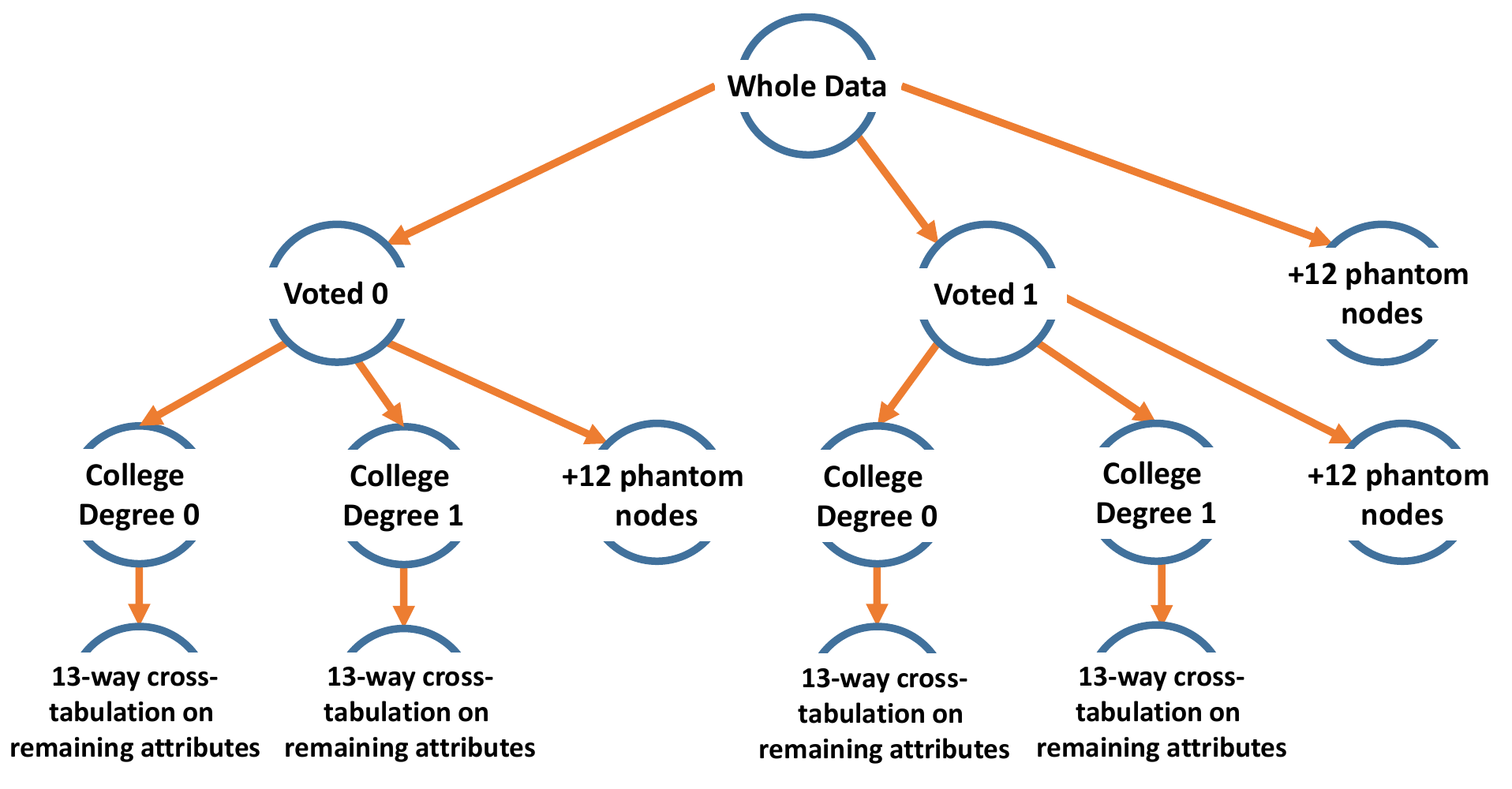}}\vspace{-6pt}
\caption{A sketch of the constructed tree in STEPS-2.}\label{fig:tree}
\end{figure}

For STEPS, we applied Algorithm \ref{alg:steps}. When deciding on the variable order for partitioning and building the tree, we leveraged domain knowledge in pubic policy instead of using the exponential mechanism.  Our domain expert, a public policy researcher at the Urban Institute, suggested  that ``Voted'', ``College Degree'', and ``Preregistration State'' are the top three important variables to practitioners in this data set. ``Voted'' is ranked first, because this is voter data set and ``voted'' is perhaps the variable that is of interest to most researchers and practitioners who use the data. Also, this attribute presents an important predictor for turnout in future elections \citep{malchow2004predicting}. 
``College Degree'' is ranked the second since scholars have consistently noted that highly educated individuals participate in politics more than the average citizen \citep{campbell1980american, rosenstone1993mobilization}. Some researchers have also concluded that education is the socio-demographic variable most strongly correlated with turnout \citep{wolfinger1980votes}, which also relates to education status. The third variable is ``Pre-registration State''. \citet{mcdonald2010registering} found higher turnout rates in Florida and Hawaii among those who preregistered (registered before they were able to take part in the next election) compared to those who registered after they turned 18. Pre-registrants were 4.7\% more likely to vote in the 2008 election than those who registered after they turned 18. For comparison, we built a tree with $L=2$ (partitioning by ``voted'' then ``College Degree''), and a tree with $L=3$ (partitioning by ``voted'' then ``College Degree'', then ``Pre-registration State''). Figure \ref{fig:tree} shows  the tree structure for STEPS at $L=2$. Since ``family income'' has the most categories, thus $b=14$ and the phantom categories are created for all the other 14 attributes to have $b=14$ universally. These phantom categories are created solely for the purposes of applying the inconsistency correction rule in  Eqns (\ref{eqn:bottom}) and (\ref{eqn:top}), and will be removed once the synthetic data are generated. 

For UH, We did not set the number of layers at $L=\log_bN$ per the original procedure \cite{hay2010boosting}, but used $L=2$ and $L=3$. The main reason is the computational constraints and practical limitations. If we had used $b=14$ with 15 attributes, with $L=15$ and $N=14^{15}$, and the computational complexity would be  $O(N)=O(14^{15})$. In each layer for $L=2$ and $L=3$, we randomly chose an attribute to partition the data, sanitized the node counts, and then applied Eqns (\ref{eqn:bottom}) and (\ref{eqn:top}) to obtain the final differentially private histogram, from which data were synthesized. 

For PrivBayes, we applied the GitHub codes \citep{DataSynthesizercodes}. The degree of the Bayesian network was set at 2  and the total $\epsilon$ was divided in half between selecting a Bayesian network vs. sanitizing the resultant joint distribution. 

The flat Laplace sanitizer injected independent noises from Laplace$(0, \epsilon^{-1})$ to each of the $1,290,240$ cells, after removing the impossible case of a person voting if they are not registered, from the  15-way cross-tabulation (Table \ref{tab:cell}). Given the enormous number of cells, the majority of the cells are empty (98.35\%). Since the empty cells are sample zeros rather than population zeros, they were and should be sanitized rather than being kept at zero. Data were synthesized from the 15-dimensional sanitized table. 

\begin{table}[!htb]\centering
\resizebox{0.8\textwidth}{!}{
\begin{tabular}{ c | c c c c c c c}
\hline
Cell size & 0 & 1 & 2 &3 & 4 & 5 & $\boldsymbol{>}$ 5\\
\hline
Number of Cells & 1,268,911 & 14,061 & 3,545 & 1,508 & 740 & 425 & 1,050 \\
\hline
Proportion & 98.35\% & 1.09\% & 0.27\% & 0.12\% & 0.06\% & 0.03\% & 0.08\% \\
\hline
\end{tabular}}
\caption{Summary of cell sizes and frequencies in the full cross tabulation of the youth voter data.}\label{tab:cell}
\end{table}

We varied privacy budget at $\epsilon\in \exp\{-2,-1,0,1,2\}$ to examine how $\epsilon$ affects the utility of the synthetic data in each of the DIPS methods. We generated $5$ synthetic data sets, each at a budget of $\epsilon/5$ to account for the synthesis and sanitization variability. We ran 10 repetitions to quantify the stability of the DIPS methods.

\subsection{Statistical Utility Assessment}\label{sec:results}
We assess the statistical utility of the synthetic data via three analyses: the SPECKS metric, chi-squared tests of association, and a difference-in-differences (DID) model.  As discussed in Sec \ref{sec:specks}, the SPECKS metric measures the worst-case separation in two data sets as a whole, whereas the latter two focus on exploring relationships among the attributes that practitioners would be interested in learning from this data set. Interested readers may refer to \citet{bowen2019comparative} for further discussion on different types of utility analysis.

In the SPECKS analysis, we used the logistic regression that contains the main effects and first order interactions among all the attributes as predictor  as the classifier. The results from the 5 synthetic data sets were averaged.  

For the chi-squared tests of association, we conducted tests for all possible 2-way tables (105 in total) across the 15 attributes to see how well the statistically significant 2-way associations detected in the original data are preserved in the synthetic data. Since the goal here is not to to pick at least one statistically significant associations out of 105 tables, we did not apply the multiplicity adjustment. We obtained the p-values from the chi-squared tests in each synthetic set and combined the p-values across the 5 synthetic sets via the combination rule in \citet{li1991significance}. 

For the third analysis, we ran the DID logistic regression in \citet{Holbein2016} to examine the effects of ``Preregistration State'' and `Registration Status'' on ``Voted''; all 14 attributes are included as predictors plus an interaction term between ``Preregistration State'' and ``Registration Status''. ``Age'' is treated  as a numerical predictor whereas the others are categorical, leading to a total of 32 regression coefficients, including the intercept. We applied the inferential combination rules in \citet{liu2016model} to obtain the CIs for the coefficients.  We calculated the CI overlap  metric \citep{karr2006framework}, which measures how much the CI estimation for a parameter based on the original data overlaps that based on the synthetic data. A value of 1 corresponds to perfect match between the CIs based on the synthetic and original data; and a value of 0 means there is no overlap between the two CIs (regardless of the degree of non-overlap). When one CI is completely contained within the other, the CI overlap value is greater than 0.5. 

\subsubsection{Utility Result Summary}

We summarize the  utility results from all three types of analysis in Table \ref{tab:utility}.  The numbers in parentheses are their performance rank, where 1 is the best and 6 is the worst. The overall score is the sum of all ranks: the smaller, the better.  The consistency rate column refers to the percentage that the chi-squared test conclusions are consistent between the original and synthetic data out of 105 tests at the significance levels of $\alpha=\{1, 5, 10\}\%$, respectively.  For the CI overlap metric, we report that the average CI overlap values across all 32 coefficients, as well as those on the coefficients associated with covariates ``Preregistration State'', ``Registration Status'', and the interaction term between ``Preregistration State'' and ``Registration Status'' (Prereg. x Reg.), which are the three covariates are of primary interest in the DID model of \citet{Holbein2016}. 
\begin{table}[!htb]
\def\arraystretch{1}
\centering\small
\resizebox{1.01\textwidth}{!}{
\begin{tabular}{ C{0.85cm} | C{2.5cm} | C{1.5cm} | C{1.5cm} | C{1.5cm} | C{1.5cm} | C{1.5cm} | C{1.5cm} | C{1.5cm} | C{1.5cm} | C{1.25cm}}
\hline
 \multirow{3}{*}{$\log(\epsilon$)} & \multirow{3}{*}{DIPS Method} & \multirow{3}{*}{\shortstack[c]{SPECKS:\\ KS \\ Distance }} &  \multicolumn{3}{c|}{Consistency Rate with $\alpha=$}  & \multicolumn{4}{c|}{Confidence Interval Overlap} & \multirow{3}{*}{\shortstack[c]{Overall \\ Score}} \\ 
\cline{4-10}
&  &  & 1\% & 5\% &  10\% & All Coeff. & Preg. State & Reg. Status & Prereg. x Reg. & \\
\hline
-2 & flat Laplace & 0.981 (6) & 0.190 (1) & 0.210 (1) & 0.210 (1) & 0.092 (3) & 0.872 (4) & 0.637 (3) & 0.872 (3) & 22 \\
-2 & PrivBayes         & 0.462 (1) & 0.152 (6) & 0.143 (6) & 0.105 (6) & 0.043 (6) & 0.681 (6) & 0.000 (6) & 0.681 (6) & 43 \\
-2 & UH modified-2     & 0.973 (4) & 0.162 (2) & 0.152 (2) & 0.114 (2) & 0.099 (1) & 0.865 (3) & 0.602 (4) & 0.865 (4) & 22 \\
-2 & UH modified-3     & 0.957 (3) & 0.162 (2) & 0.152 (2) & 0.114 (2) & 0.093 (2) & 0.848 (5) & 0.495 (5) & 0.848 (5) & 26 \\
-2 & STEPS-2           & 0.975 (5) & 0.162 (2) & 0.152 (2) & 0.114 (2) & 0.090 (4) & 0.874 (1) & 0.661 (1) & 0.874 (1) & 18 \\
-2 & STEPS-3           & 0.956 (2) & 0.162 (2) & 0.152 (2) & 0.114 (2) & 0.090 (4) & 0.873 (2) & 0.659 (2) & 0.873 (2) & 18 \\
\hline

-1 & flat Laplace & 0.978 (6) & 0.181 (1) & 0.210 (1) & 0.210 (1) & 0.102 (1) & 0.871 (3) & 0.638 (3) & 0.871 (3) & 19\\
-1 & PrivBayes         & 0.422 (1) & 0.152 (6) & 0.143 (6) & 0.105 (6) & 0.050 (6) & 0.681 (6) & 0.000 (6) & 0.681 (6) & 43\\
-1 & UH modified-2     & 0.971 (4) & 0.162 (2) & 0.152 (4) & 0.114 (4) & 0.098 (2) & 0.865 (4) & 0.602 (4) & 0.865 (4) & 28\\
-1 & UH modified-3     & 0.954 (2) & 0.162 (2) & 0.152 (4) & 0.114 (4) & 0.092 (3) & 0.839 (5) & 0.445 (5) & 0.839 (5) & 30\\
-1 & STEPS-2           & 0.971 (4) & 0.162 (2) & 0.171 (2) & 0.133 (2) & 0.090 (5) & 0.874 (1) & 0.667 (1) & 0.874 (1) & 18 \\
-1 & STEPS-3           & 0.954 (2) & 0.162 (2) & 0.171 (2) & 0.133 (2) & 0.091 (4) & 0.874 (1) & 0.666 (2) & 0.874 (1) & 16\\
\hline

0 & flat Laplace & 0.968 (6) & 0.190 (1) & 0.210 (1) & 0.210 (1) & 0.108 (1) & 0.871 (4) & 0.641 (3) & 0.871 (4) & 21\\
0 & PrivBayes         & 0.272 (1) & 0.152 (6) & 0.143 (6) & 0.114 (4) & 0.076 (6) & 0.943 (1) & 0.000 (6) & 0.962 (1) & 31\\
0 & UH modified-2     & 0.964 (4) & 0.162 (4) & 0.152 (4) & 0.114 (4) & 0.106 (2) & 0.864 (5) & 0.605 (4) & 0.864 (5) & 32 \\
0 & UH modified-3     & 0.945 (2) & 0.162 (4) & 0.152 (4) & 0.114 (4) & 0.090 (3) & 0.838 (6) & 0.446 (5) & 0.838 (6) & 34\\
0 & STEPS-2           & 0.965 (5) & 0.181 (2) & 0.171 (2) & 0.133 (2) & 0.090 (3) & 0.873 (2) & 0.671 (1) & 0.873 (2) & 19 \\
0 & STEPS-3           & 0.945 (2) & 0.181 (2) & 0.171 (2) & 0.133 (2) & 0.090 (3) & 0.873 (2) & 0.671 (1) & 0.873 (2) & 16 \\
\hline

1 & flat Laplace & 0.930 (4) & 0.276 (1) & 0.371 (1) & 0.448 (1) & 0.092 (4) & 0.867 (4) & 0.651 (4) & 0.867 (4) & 23 \\
1 & PrivBayes         & 0.187 (1) & 0.171 (4) & 0.171 (6) & 0.133 (6) & 0.102 (2) & 0.978 (1) & 0.847 (1) & 0.990 (1) & 22\\
1 & UH modified-2     & 0.944 (5) & 0.162 (5) & 0.200 (4) & 0.219 (4) & 0.125 (1) & 0.863 (5) & 0.612 (5) & 0.863 (5) & 34 \\
1 & UH modified-3     & 0.919 (2) & 0.162 (5) & 0.181 (5) & 0.219 (4) & 0.096 (3) & 0.837 (6) & 0.453 (6) & 0.837 (6) & 37 \\
1 & STEPS-2           & 0.944 (5) & 0.200 (2) & 0.257 (2) & 0.238 (2) & 0.091 (5) & 0.871 (2) & 0.677 (2) & 0.871 (2) & 22 \\
1 & STEPS-3           & 0.919 (2) & 0.200 (2) & 0.248 (2) & 0.238 (2) & 0.084 (6) & 0.871 (2) & 0.676 (3) & 0.871 (2) & 21 \\
\hline

2 & flat Laplace & 0.833 (2) & 0.467 (1) & 0.552 (1) & 0.629 (1) & 0.104 (1) & 0.862 (4) & 0.684 (4) & 0.862 (4) & 18 \\
2 & PrivBayes         & 0.157 (1) & 0.190 (6) & 0.181 (6) & 0.143 (7) & 0.099 (3) & 0.988 (1) & 0.983 (1) & 0.971 (1) & 26\\
2 & UH modified-2     & 0.885 (5) & 0.352 (4) & 0.400 (4) & 0.410 (6) & 0.100 (2) & 0.860 (5) & 0.635 (5) & 0.860 (5) & 36\\
2 & UH modified-3     & 0.843 (3) & 0.305 (5) & 0.390 (5) & 0.457 (4) & 0.076 (5) & 0.849 (6) & 0.553 (6) & 0.849 (6) & 40\\
2 & STEPS-2           & 0.886 (6) & 0.400 (2) & 0.448 (2) & 0.486 (2) & 0.093 (4) & 0.865 (2) & 0.696 (2) & 0.865 (2) & 22\\
2 & STEPS-3           & 0.843 (3) & 0.400 (2) & 0.448 (2) & 0.486 (2) & 0.076 (5) & 0.865 (2) & 0.695 (3) & 0.865 (2) & 21\\   
\hline 
\end{tabular}}
\caption{Summary utility analysis. The numbers in parentheses are the performance ranks with 1 being the best and 6 being the worst. Overall score is the sum of all ranks: the smaller, the better.}\label{tab:utility}
\end{table}

Overall, the utility in each analysis increases with $\epsilon$ for each DIP method, with the only exception on the CI overlap values from the DID model, where the improvement with $\epsilon$ is not obvious. All analyses taken together, STEPS-2 and STEPS-3 with $L=2$ and $L=3$ are the best, followed by the flat Laplace sanitizer. PrivBayes performs the best with regard to the SPECKS analysis, by a large margin, compared to the rest, but does not do well in the chi-squared tests or the DID model CI overlap analysis. The Laplace sanitizer performs the best in the chi-squared tests, slightly worse than STEPS on the other analysis especially when $\epsilon$ is small. The STEPS method outperforms the random partitioning approach via the modified UH method on almost all metrics, demonstrating the advantages of informed partitioning.



In what follows, we provide more details on the results in each of the utility analyses.

\subsubsection{SPECKS Analysis for General Utility}\label{sec:specks-results}
\begin{figure}[!htb]
\centerline{\includegraphics[scale=0.45]{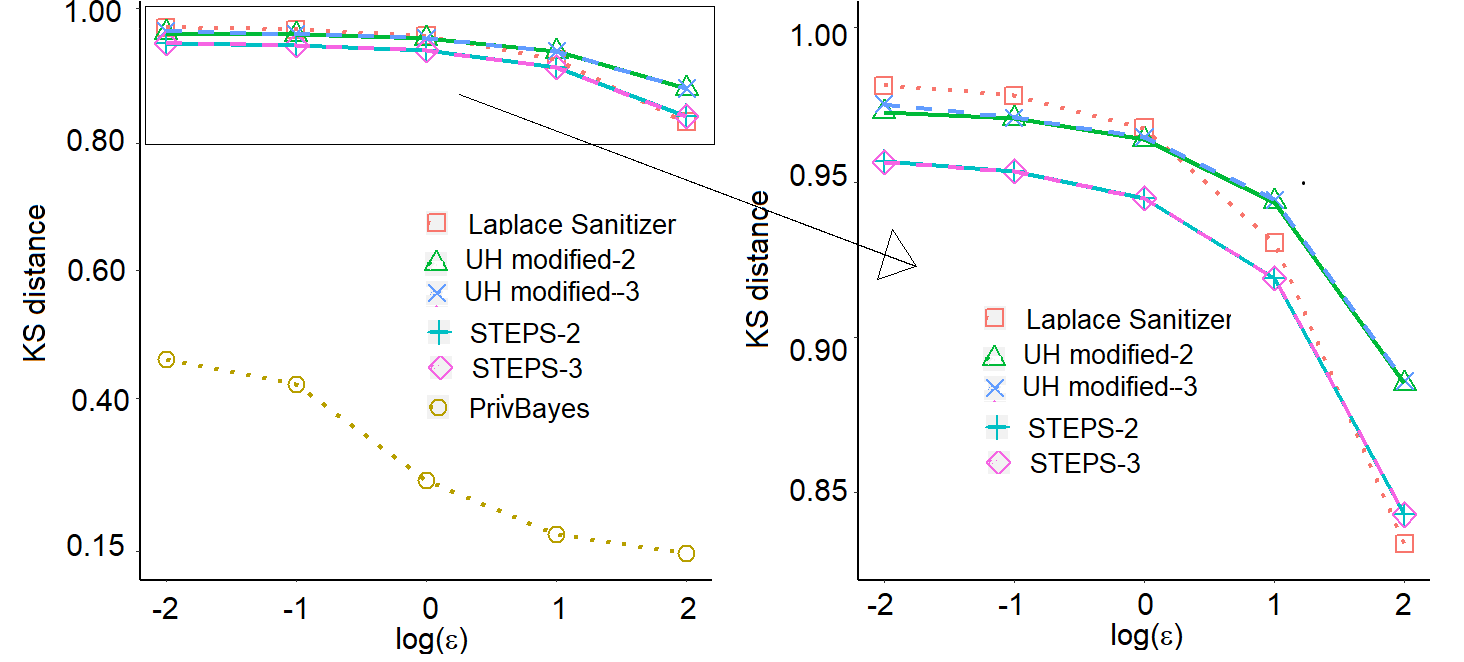}}
\vspace{-6pt}\caption{SPECKS analysis. The plot on the left presents all the methods and the one on the right is a zoom-in on the boxed area.} \label{fig:SPECKS}
\end{figure}

Figure \ref{fig:SPECKS} depicts the results on the SPECKS analysis   (the standard deviations on the KS distance across the repeats are two orders of magnitude smaller than the average KS distance and barely visible on the plots). For all methods, the KS distance decreases as $\epsilon$ increases, as expected, providing empirical evidence that SPECKS does what it is supposed to measure. Overall, PrivBayes outperforms all of the other methods for all levels of $\epsilon$. STEPS-2 and STEPS-3 are very similar and slightly outperform the flat Laplace sanitizer (until $\epsilon=e^2$) and the UH modified. 

\subsubsection{Chi-squared Tests of Association}\label{sec:chi}
\begin{figure}[!htb]
\centerline{\includegraphics[width=6.5in]{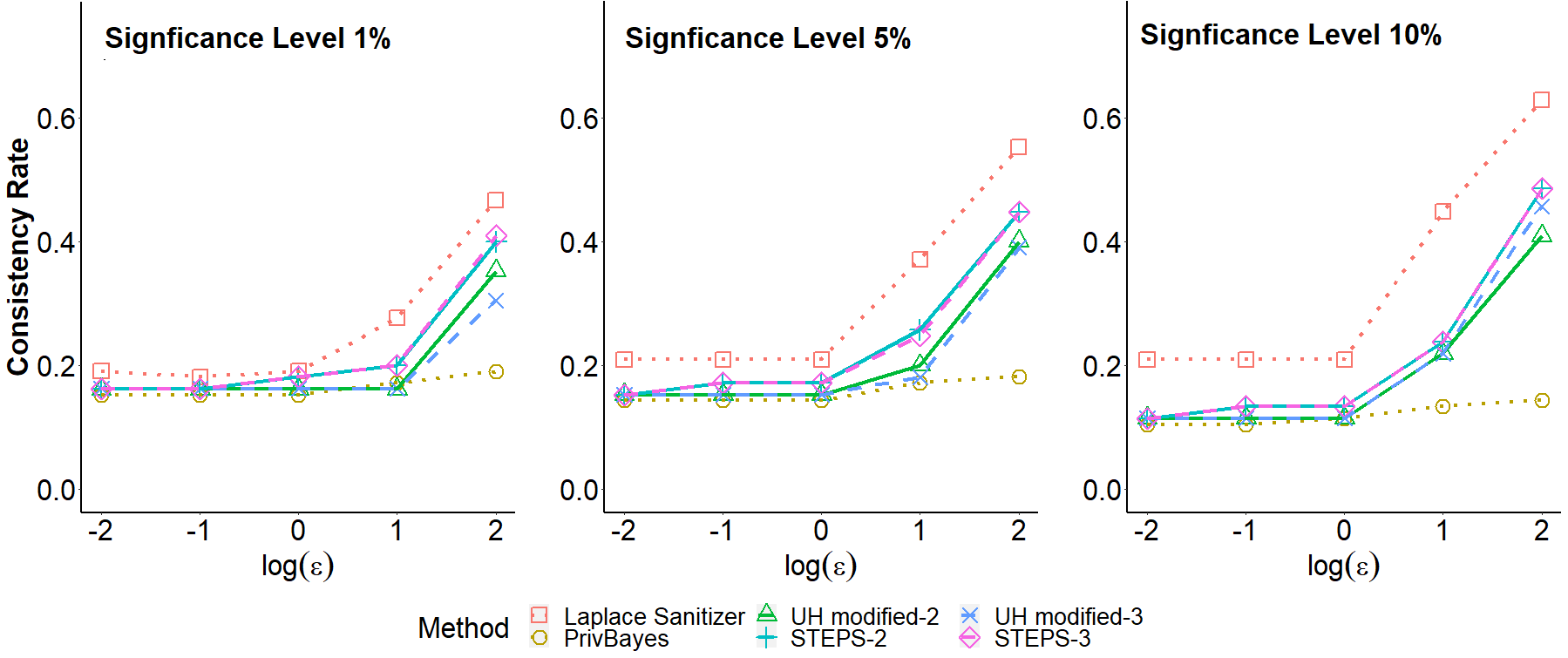}}\vspace{-6pt}
\caption{Consistency rate on the statistical significance in the 105 Chi-square tests of association based on the original and the synthetic data.} \label{fig:chisq}\vspace{-6pt}
\end{figure}

Figure \ref{fig:chisq} presents the chi-squared test statistical significance consistency rates at $\alpha=$ 1\%, 5\%, 10\%, respectively. For all values of $\epsilon$, the Laplace sanitizer performs the best, followed by the STEPS methods. The UH modified methods perform slightly worse than the STEPS methods while PrivBayes remains at  consistently low rates across all levels of $\epsilon$. We also present in the Appendix Tables \ref{tab:lap} to \ref{tab:step3} that 
contain the cross-tabulations of the original vs. sanitized p-value by categories $\le0.01$, $(0.01,0.05]$, $(0.05,0.1]$, and $>0.1$. The tables suggest there is improvement in the  alignment between the original and sanitized p-values as $\epsilon$ increases, but in a less satisfactory manner with this more granular categorization, even for $\epsilon$ as large as $e^2\approx7.4$, compared to the binary classification on the p-values plotted in Figure \ref{fig:chisq}. 

\subsubsection{Difference-in-Differences (DID) Model}\label{sec:did}
\begin{figure}[!htb]
\centerline{\includegraphics[width=5in]{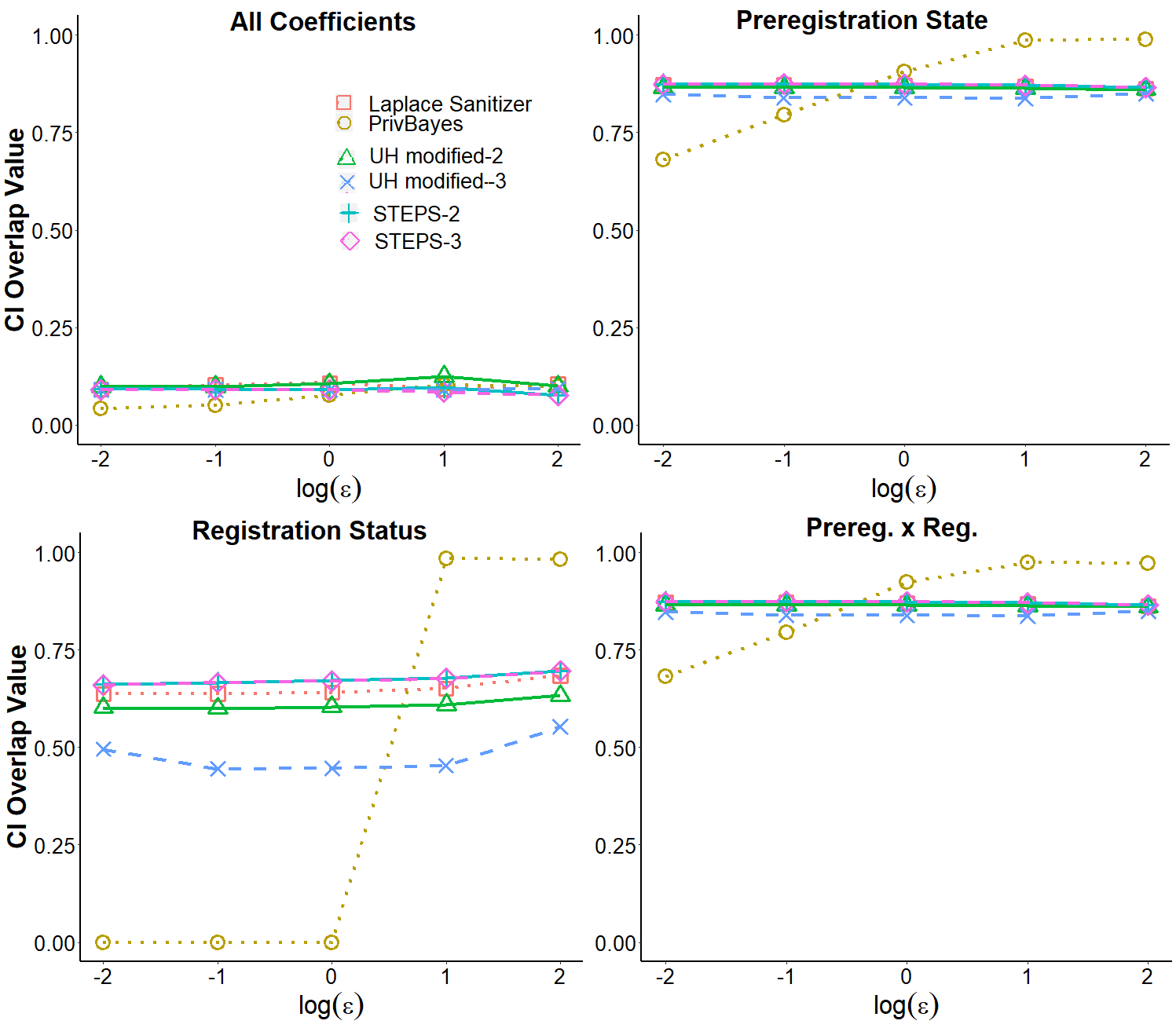}}
\vspace{-6pt}\caption{Confidence interval (CI) overlap value analysis averaged across the 32 regression coefficients from the DID model fitted to the synthetic data generated from the flat Laplace sanitizer, PrivBayes, STEPS, and a modified UH with random partition.} \label{fig:DID}
\end{figure}

Figure \ref{fig:DID} shows the results for the average CIs across all 32 regression coefficients, that on the coefficients for ``Preregistration State'', ``Registration Status'', and the interaction term between ``Preregistration State'' and ``Registration Status'' (Prereg. x Reg.), respectively form the DID model. The effects of the latter three covariates are of primary interest in \citet{Holbein2016}. 

For the average CI overlap value across the 32 coefficients, since a non-trivial proportion of the CI overlap values are close 0, the averaged CI is close to 0 and is not very informative in  differentiating the methods. For the three covariates of interest, the CI overlap values are around 0.75 or larger in most cases, suggesting good agreement between the CIs based on the synthetic and  original data. Among the four DIPS methods, PrivBayes has the lowest CI overlap values when $\log(\epsilon) < 0$, and performs better as $\epsilon$ increase. The CI overlap values are similar for all the other methods and across the examined $\epsilon$ values.

\section{Concluding Remarks}\label{sec:disc}
We propose the STEPS procedure to synthesize differentially private individual-level data. We also develop a propensity-based general utility metric, SPECKS, for assessing the similarity between the original and synthetic data. The STEPS method leverages the inherent information in the data or public domain knowledge to build a hierarchical tree among the attributes. The sanitized counts of the nodes closer to the root of the tree (low-order marginals) have smaller mean squared errors than the nodes further away from the root because the former are weighted averages of multiple sources of information. STEPS therefore preserves the information on more important attributes better than on less important ones, placing those attributes closer to the root of the tree. 

We used the PrivBayes codes on GitHub at \citet{DataSynthesizercodes} developed by \citet{DataSynthesizer}. We noticed during the implementation that we needed to set a random seed in each synthetic generate; otherwise, every synthetic set is the same for a given $\epsilon$. Users of the codes need to keep that in mind when generating multiple synthetic sets to measure the synthesis uncertainty. This is just one example on the importance of good documentation when publishing and sharing nodes for implementing differentially private mechanisms and data synthesis methods to avoid misleading interpretation or results. \citet{kifer2020guidelines} 
encourage that every DP mechanism provides an accompanying mathematical proof on DP (not a just reference to the literature) along with proper code review and computational-aided verification/testing tools. 
We will publish the codes used in the simulation and cases studies in this paper on GitHub to facilitate the practical implementation of the proposed methods and for reproducibility checks. 

For future work, we will look into developing general recommendations on the privacy budget allocation scheme between building a tree and sanitizing node counts  among the tree layers. In addition, we will investigate the feasibility of developing a stopping rule on the tree height, from the utility and computational prospective, instead of pre-specifying the number of layers. 
Finally,  we plan to compare SPECKS with other propensity-score based general utility measures via comprehensive empirical studies.

\section*{Acknowledgement}
We thank Madeline Brown, a Policy Assistant, at the Urban Institute for her expert knowledge in selecting variable order as a domain expert in the case study. We also thank the associate editor and two referees who provided useful feedback to improve the manuscript.

\bibliographystyle{abbrvnat}
\setlength{\bibhang}{0pt}
\bibliography{reflist}

\newpage
\section*{Appendix}

\begin{table}[ht]
\def\arraystretch{1.0}
\centering
\begin{tabular}{c|rrrrr|c}
  \hline
        & \multicolumn{5}{c|}{sanitized} & consistency\\
  \cline{2-6}
        $\log(\epsilon) $ & Original & $< 0.01$ & $0.01-0.05$ & $0.05-0.1$ & $>0.1$  & count$^\ddagger$ \\ 
  \hline
       \multirow{4}{*}{-2} & $< 0.01$ &   5 &   5 &   0 &  79  & \multirow{4}{*}{15} \\ 
         & $0.01-0.05$ &   0 &   0 &   0 &   1 \\ 
         & $0.05-0.1$ &   0 &   2 &   0 &   2 \\ 
         & $>0.1$ &   1 &   0 &   0 &  10 \\ 
  \hline
        \multirow{4}{*}{-1} & $< 0.01$ &  4 &   6 &   0 &  79 & \multirow{4}{*}{14}\\ 
         & $0.01-0.05$ &   0 &   0 &   0 &   1 \\ 
         & $0.05-0.1$ &  0 &   2 &   0 &   2 \\ 
         & $>0.1$ &  1 &   0 &   0 &  10 \\ 
  \hline
        \multirow{4}{*}{0} & $< 0.01$ &    5 &   5 &   0 &  79 & \multirow{4}{*}{15} \\ 
         & $0.01-0.05$ &   0 &   0 &   0 &   1 \\ 
         & $0.05-0.1$ &  0 &   2 &   0 &   2 \\ 
         & $>0.1$ &  1 &   0 &   0 &  10 \\ 
  \hline
        \multirow{4}{*}{1} & $< 0.01$ &  19 &  12 &   4 &  54 & \multirow{4}{*}{27} \\  
         & $0.01-0.05$ &   0 &   0 &   0 &   1 \\ 
         & $0.05-0.1$ &  3 &   1 &   0 &   0 \\ 
         & $>0.1$ &  3 &   0 &   0 &   8 \\ 
  \hline
        \multirow{4}{*}{2} & $< 0.01$ &  41 &  11 &   6 &  31  & \multirow{4}{*}{45} \\ 
         & $0.01-0.05$ &   0 &   0 &   0 &   1 \\ 
         & $0.05-0.1$ &  4 &   0 &   0 &   0 \\ 
         & $>0.1$ &  4 &   1 &   2 &   4 \\ 
   \hline
\end{tabular}
\begin{tabular}{l}
\footnotesize $^\ddagger$ sum of diagonal counts.\hspace{3.6in}\textcolor{white}.\\
 \hline
\end{tabular}
\caption{The cross-tabulation of the 105 p-values by categories $<0.01$, $0.01-0.05$, $0.05-0.1$, and $>0.1$ for the original and DIPS data generated by the Laplace sanitizer.} \label{tab:lap}
\end{table}

\begin{table}[ht]
\def\arraystretch{1.1}
\centering
\begin{tabular}{c|rrrrr|c}
  \hline
        & \multicolumn{5}{c|}{sanitized} & consistency\\
  \cline{2-6}
        $\log(\epsilon) $ & Original & $< 0.01$ & $0.01-0.05$ & $0.05-0.1$ & $>0.1$  & count$^\ddagger$ \\ 
  \hline
       \multirow{4}{*}{-2} & $< 0.01$ &  0 &   0 &   0 &  89 & \multirow{4}{*}{11}\\ 
         & $0.01-0.05$ &  0 &   0 &   0 &   1 \\ 
         & $0.05-0.1$ &   0 &   0 &   0 &   4 \\ 
         & $>0.1$ &   0 &   0 &   0 &  11 \\ 
  \hline
        \multirow{4}{*}{-1} & $< 0.01$ &   0 &   0 &   0 &  89 & \multirow{4}{*}{11} \\ 
         & $0.01-0.05$ &   0 &   0 &   0 &   1 \\
         & $0.05-0.1$ &  0 &   0 &   0 &   4 \\ 
         & $>0.1$ &  0 &   0 &   0 &  11 \\ 
  \hline
        \multirow{4}{*}{0} & $< 0.01$ &   0 &   0 &   1 &  88 & \multirow{4}{*}{11}\\ 
         & $0.01-0.05$ &   0 &   0 &   0 &   1 \\ 
         & $0.05-0.1$ &   0 &   0 &   0 &   4 \\  
         & $>0.1$ &  0 &   0 &   0 &  11 \\ 
  \hline
        \multirow{4}{*}{1} & $< 0.01$ &   2 &   1 &   0 &  86 & \multirow{4}{*}{13}\\ 
         & $0.01-0.05$ &  0 &   0 &   0 &   1 \\ 
         & $0.05-0.1$ &  0 &   0 &   0 &   4 \\ 
         & $>0.1$ &  0 &   0 &   0 &  11 \\ 
  \hline
        \multirow{4}{*}{2} & $< 0.01$ &   4 &   0 &   0 &  85 & \multirow{4}{*}{15}\\ 
         & $0.01-0.05$ &   0 &   0 &   0 &   1 \\ 
         & $0.05-0.1$ &   0 &   0 &   0 &   4 \\ 
         & $>0.1$ &   0 &   0 &   0 &  11 \\ 
   \hline
\end{tabular}
\begin{tabular}{l}
\footnotesize $^\ddagger$ sum of diagonal counts.\hspace{3.6in}\textcolor{white}.\\
 \hline
\end{tabular}
\caption{The cross-tabulation of the 105 p-values in categories $<0.01$, $0.01-0.05$, $0.05-0.1$, and $>0.1$ for the original and DIPS data generated by the PrivBayes.} \label{tab:priv}
\end{table}

\begin{table}[ht]
\def\arraystretch{1.1}
\centering
\begin{tabular}{c|rrrrr|c}
  \hline
        & \multicolumn{5}{c|}{sanitized} & consistency\\
  \cline{2-6}
        $\log(\epsilon) $ & Original & $< 0.01$ & $0.01-0.05$ & $0.05-0.1$ & $>0.1$  & count$^\ddagger$ \\ 
  \hline
       \multirow{4}{*}{-2} & $< 0.01$ &  1 &   0 &   0 &  88 & \multirow{4}{*}{12}\\ 
         & $0.01-0.05$ &  0 &   0 &   0 &   1 \\ 
         & $0.05-0.1$ &   0 &   0 &   0 &   4 \\ 
         & $>0.1$ &   0 &   0 &   0 &  11 \\ 
  \hline
        \multirow{4}{*}{-1} & $< 0.01$ &   1 &   0 &   0 &  88 & \multirow{4}{*}{12}\\ 
         & $0.01-0.05$ &   0 &   0 &   0 &   1 \\ 
         & $0.05-0.1$ &   0 &   0 &   0 &   4 \\ 
         & $>0.1$ &   0 &   0 &   0 &  11 \\ 
  \hline
        \multirow{4}{*}{0} & $< 0.01$ &  1 &   0 &   0 &  88 & \multirow{4}{*}{12}\\ 
         & $0.01-0.05$ &   0 &   0 &   0 &   1 \\ 
         & $0.05-0.1$ &   0 &   0 &   0 &   4 \\  
         & $>0.1$ &   0 &   0 &   0 &  11 \\ 
  \hline
        \multirow{4}{*}{1} & $< 0.01$ &  1 &   7 &   4 &  77 & \multirow{4}{*}{10}\\ 
         & $0.01-0.05$ &   0 &   0 &   0 &   1 \\ 
         & $0.05-0.1$ &   0 &   2 &   0 &   2 \\ 
         & $>0.1$ &   0 &   0 &   2 &   9 \\ 
  \hline
        \multirow{4}{*}{2} & $< 0.01$ &  22 &   8 &   5 &  54 & \multirow{4}{*}{29}\\ 
         & $0.01-0.05$ &   0 &   0 &   0 &   1 \\ 
         & $0.05-0.1$ &   0 &   0 &   1 &   3 \\ 
         & $>0.1$ &   1 &   2 &   1 &   7 \\
   \hline
\end{tabular}
\begin{tabular}{l}
\footnotesize $^\ddagger$ sum of diagonal counts.\hspace{3.6in}\textcolor{white}.\\
 \hline
\end{tabular}
\caption{The cross-tabulation of the 105 p-values by categories $<0.01$, $0.01-0.05$, $0.05-0.1$, and $>0.1$ for the original and DIPS data generated by the UH Modified-2.} \label{tab:uh2}
\end{table}

\begin{table}[ht]
\def\arraystretch{1.1}
\centering
\begin{tabular}{c|rrrrr|c}
  \hline
        & \multicolumn{5}{c|}{sanitized} & consistency\\
  \cline{2-6}
        $\log(\epsilon) $ & Original & $< 0.01$ & $0.01-0.05$ & $0.05-0.1$ & $>0.1$  & count$^\ddagger$ \\ 
  \hline
       \multirow{4}{*}{-2} & $< 0.01$ &   1 &   0 &   0 &  88 & \multirow{4}{*}{12}\\ 
         & $0.01-0.05$ &   0 &   0 &   0 &   1 \\ 
         & $0.05-0.1$ &   0 &   0 &   0 &   4 \\ 
         & $>0.1$ &   0 &   0 &   0 &  11 \\ 
  \hline
        \multirow{4}{*}{-1} & $< 0.01$ &   1 &   0 &   0 &  88 & \multirow{4}{*}{12}\\ 
         & $0.01-0.05$ &   0 &   0 &   0 &   1 \\ 
         & $0.05-0.1$ &   0 &   0 &   0 &   4 \\ 
         & $>0.1$ &   0 &   0 &   0 &  11 \\ 
  \hline
        \multirow{4}{*}{0} & $< 0.01$ &   1 &   0 &   0 &  88 & \multirow{4}{*}{12}\\ 
         & $0.01-0.05$ &   0 &   0 &   0 &   1 \\ 
         & $0.05-0.1$ &   0 &   0 &   0 &   4 \\ 
         & $>0.1$ &   0 &   0 &   0 &  11 \\ 
  \hline
        \multirow{4}{*}{1} & $< 0.01$ &   1 &   6 &   5 &  77 & \multirow{4}{*}{10}\\ 
         & $0.01-0.05$ &   0 &   0 &   0 &   1 \\ 
         & $0.05-0.1$ &   0 &   2 &   0 &   2 \\ 
         & $>0.1$ &   0 &   1 &   1 &   9 \\ 
  \hline
        \multirow{4}{*}{2} & $< 0.01$ &  20 &  13 &   8 &  48 & \multirow{4}{*}{25}\\ 
         & $0.01-0.05$ &   0 &   0 &   0 &   1 \\ 
         & $0.05-0.1$ &   1 &   1 &   0 &   2 \\ 
         & $>0.1$ &   3 &   2 &   1 &   5 \\ 
   \hline
\end{tabular}
\begin{tabular}{l}
\footnotesize $^\ddagger$ sum of diagonal counts.\hspace{3.6in}\textcolor{white}.\\
 \hline
\end{tabular}
\caption{The cross-tabulation of the 105 p-values by categories $<0.01$, $0.01-0.05$, $0.05-0.1$, and $>0.1$ for the original and DIPS data generated by the UH Modified-3.} \label{tab:uh3}
\end{table}

\begin{table}[ht]
\def\arraystretch{1.1}
\centering
\begin{tabular}{c|rrrrr|c}
  \hline
        & \multicolumn{5}{c|}{sanitized} & consistency\\
  \cline{2-6}
        $\log(\epsilon) $ & Original & $< 0.01$ & $0.01-0.05$ & $0.05-0.1$ & $>0.1$  & count$^\ddagger$ \\ 
  \hline
       \multirow{4}{*}{-2} & $< 0.01$ &   1 &   0 &   0 &  88 & \multirow{4}{*}{12}\\ 
         & $0.01-0.05$ &   0 &   0 &   0 &   1 \\ 
         & $0.05-0.1$ &   0 &   0 &   0 &   4 \\ 
         & $>0.1$ &   0 &   0 &   0 &  11 \\ 
  \hline
        \multirow{4}{*}{-1} & $< 0.01$ &    1 &   2 &   0 &  86 & \multirow{4}{*}{12}\\ 
         & $0.01-0.05$ &   0 &   0 &   0 &   1 \\ 
         & $0.05-0.1$ &   0 &   0 &   0 &   4 \\ 
         & $>0.1$ &  0 &   0 &   0 &  11 \\ 
  \hline
        \multirow{4}{*}{0} & $< 0.01$ &  3 &   0 &   0 &  86 & \multirow{4}{*}{14}\\ 
         & $0.01-0.05$ &   0 &   0 &   0 &   1 \\ 
         & $0.05-0.1$ &   0 &   0 &   0 &   4 \\ 
         & $>0.1$ &    0 &   0 &   0 &  11 \\ 
  \hline
        \multirow{4}{*}{1} & $< 0.01$ &   5 &   8 &   4 &  72 & \multirow{4}{*}{13}\\ 
         & $0.01-0.05$ &   0 &   0 &   0 &   1 \\ 
         & $0.05-0.1$ &   0 &   0 &   0 &   4 \\ 
         & $>0.1$ &  0 &   1 &   2 &   8 \\ 
  \hline
        \multirow{4}{*}{2} & $< 0.01$ &  30 &   7 &   8 &  44 & \multirow{4}{*}{36}\\ 
         & $0.01-0.05$ &  0 &   0 &   0 &   1 \\ 
         & $0.05-0.1$ &   0 &   0 &   0 &   4 \\ 
         & $>0.1$ &   4 &   1 &   0 &   6 \\ 
   \hline
\end{tabular}
\begin{tabular}{l}
\footnotesize $^\ddagger$ sum of diagonal counts.\hspace{3.6in}\textcolor{white}.\\
 \hline
\end{tabular}
\caption{The cross-tabulation of the 105 p-values by categories $<0.01$, $0.01-0.05$, $0.05-0.1$, and $>0.1$ for the original and DIPS data generated by the STEPS-2.} \label{tab:step2}
\end{table}
\begin{table}[ht]
\def\arraystretch{1.1}
\centering
\begin{tabular}{c|rrrrr|c}
  \hline
        & \multicolumn{5}{c|}{sanitized} & consistency\\
  \cline{2-6}
        $\log(\epsilon) $ & Original & $< 0.01$ & $0.01-0.05$ & $0.05-0.1$ & $>0.1$  & count$^\ddagger$ \\ 
  \hline
       \multirow{4}{*}{-2} & $< 0.01$ &  1 &   0 &   0 &  88 & \multirow{4}{*}{12}\\ 
         & $0.01-0.05$ &   0 &   0 &   0 &   1 \\ 
         & $0.05-0.1$ &   0 &   0 &   0 &   4 \\ 
         & $>0.1$ &   0 &   0 &   0 &  11 \\ 
  \hline
        \multirow{4}{*}{-1} & $< 0.01$ &   1 &   2 &   0 &  86 & \multirow{4}{*}{12}\\ 
         & $0.01-0.05$ &   0 &   0 &   0 &   1 \\ 
         & $0.05-0.1$ &   0 &   0 &   0 &   4 \\ 
         & $>0.1$ &   0 &   0 &   0 &  11 \\ 
  \hline
        \multirow{4}{*}{0} & $< 0.01$ &   3 &   0 &   0 &  86 & \multirow{4}{*}{14}\\ 
         & $0.01-0.05$ &   0 &   0 &   0 &   1 \\ 
         & $0.05-0.1$ &   0 &   0 &   0 &   4 \\ 
         & $>0.1$ &  0 &   0 &   0 &  11 \\ 
  \hline
        \multirow{4}{*}{1} & $< 0.01$ &   5 &   8 &   3 &  73 & \multirow{4}{*}{14}\\ 
         & $0.01-0.05$ &   0 &   0 &   0 &   1 \\ 
         & $0.05-0.1$ &  0 &   0 &   0 &   4 \\ 
         & $>0.1$ &   0 &   2 &   0 &   9 \\ 
  \hline
        \multirow{4}{*}{2} & $< 0.01$ &  30 &   7 &   8 &  44 & \multirow{4}{*}{36}\\
         & $0.01-0.05$ &   0 &   0 &   0 &   1 \\
         & $0.05-0.1$ &   0 &   0 &   0 &   4 \\
         & $>0.1$ &   3 &   2 &   0 &   6 \\ 
   \hline
\end{tabular}
\begin{tabular}{l}
\footnotesize $^\ddagger$ sum of diagonal counts.\hspace{3.6in}\textcolor{white}.\\
 \hline
\end{tabular}
\caption{The cross-tabulation on the 105 p-values by categories $<0.01$, $0.01-0.05$, $0.05-0.1$, and $>0.1$ for the original and DIPS data generated by the STEPS-3.} \label{tab:step3}
\end{table}

\end{document}